\documentclass[twocolumn,tighten,times]{aastex62}
\usepackage{csquotes}
\usepackage{graphicx}
\usepackage[caption=false]{subfig}
\usepackage{longtable}
\usepackage{hyperref}
\submitjournal{ApJ}

\begin{document}
\title[V1280 Sco]{PHOTOIONIZATION MODELING OF THE DUSTY NOVA V1280 SCORPII}

\correspondingauthor{Ruchi Pandey}
\email{ruchi.pandey@bose.res.in}
\author[0000-0002-6222-3045]{Ruchi Pandey}
\affiliation{S. N. Bose National Centre for Basic Sciences, Salt Lake, Kolkata 700106, India}

\author{Ramkrishna Das}
\affiliation{S. N. Bose National Centre for Basic Sciences, Salt Lake, Kolkata 700106, India}

\author{Gargi Shaw}
\affiliation{Department of Astronomy and Astrophysics, Tata Institute of Fundamental Research, Homi Bhabha Road,  Mumbai 4000005, India}

\author{Soumen Mondal}
\affiliation{S. N. Bose National Centre for Basic Sciences, Salt Lake, Kolkata 700106, India}

\label{firstpage}
\begin{abstract}
We perform photoionization modeling of the dusty nova V1280 Scorpii (V1280 Sco) with an aim to study the changes in the physical and chemical parameters. We model pre and post dust phase, optical and near-Infrared (NIR), spectra  using the  photoionization code \textsc{cloudy}, v.17.02, considering a two-component (low density and high density region) model. From the best-fit model, we find that the temperature and luminosity of the central ionizing source in the pre-dust phase is in the range 1.32 - 1.50 $\times 10^4$ K and 2.95 - 3.16 $\times 10^{36}$ ergs$^{-1}$, respectively, which increase to 1.58 - 1.62 $\times 10^4$ K and 3.23 - 3.31 $\times 10^{36}$ ergs$^{-1}$, respectively, in the post-dust phase. It is found that a very high hydrogen density ($\sim 10^{13} - 10^{14}$ cm$^{-3}$) is required for the generation of spectra properly. Dust condensation conditions are achieved at high ejecta density ($\sim 3.16 \times 10^{8}$cm$^{-3}$) and low temperature ($\sim$2000 K) in the outer region of the ejecta. It is found that a mixture of small (0.005 - 0.25$\mu$m) amorphous carbon dust grains and large (0.03 - 3.0$\mu$m) astrophysical silicate dust grains iis present n the ejecta in the post-dust phase. Our model yields a very high elemental abundance values as C/H = 13.5 - 20, N/H = 250, O/H = 27 - 35, by number, relative to solar in the ejecta, during the pre-dust phase, which decrease in the post-dust phase. 
\end{abstract}

\keywords{stars : novae, cataclysmic variables; methods: observational; techniques: spectroscopic methods: numerical, photoionization}

\section{Introduction}\label{introduction}
A \enquote{nova} refers to a close binary system that consists of a white dwarf (WD) primary and a secondary star. The white dwarf is generally a carbon-oxygen (CO) or an oxygen-neon (ONe) type, whereas, the secondary has generally been observed to be a K or M main-sequence star (Jos\'e 2012). The secondary fills its Roche lobe (Crawford \& Kraft 1956) and transfers hydrogen-rich matter through the inner Lagrangian point. The transferred matter is accreted by the primary and consequently, an accretion disk is formed around the WD and forms a layer of nuclear fuel on it. Over a period of time, the bottom of this layer gets compressed by the surface gravity of the WD and becomes electron degenerate. When the temperature at the bottom of this degenerate layer reaches the Fermi temperature ($\sim 7 \times 10^7$ K), the degeneracy becomes unimportant and the layers expand with a  further rapid increase in the temperature, and a thermonuclear runaway (TNR) ensues resulting in a nova outburst. As a result, the temperature in the nuclear burning region reaches $\sim 4 \times 10^8$ K which gives rise to significant energy production by transforming CNO nuclei to the $\beta^{+}$ -unstable nuclei like $^{13}$N, $^{14}$O, $^{15}$O, and $^{15}$F, via proton capturing. The energy released from $\beta^+$ -decay helps in the ejection of the material from the WD and also produces non-solar CNO isotopic abundances in the ejecta (Townsley \& Bildsten 2004; Yaron et al. 2005; Jos\'e \& Hernanz 2007;  Jos\'e 2012; Starrfield et al. 2008, 2012, 2016). The explosion releases significant amount of energy ($\sim 10^{40} - 10^{45} erg s^{-1} $) and ejects a shell of mass ($\sim 10^{-4} - 10^{-6}$ M$_\odot$) (Gehrz et al. 1998) with velocities ranging from hundreds to thousands of km s$^{-1}$ (Bode \& Evans 2008, and references therein).

As the ejecta expands, it becomes transparent to high energy radiation emitted by the central engine (Gallagher \& Code 1974; Balman et al. 1998, Orio et al. 2002). Hence, the novae ejecta does not appear as a suitable place for dust formation, as the hard emission from the central ionizing source does not allow the chemical routes that lead to the formation of dust grains via the formation of diatomic and polyatomic molecules (Evans \& Rawlings 1994). However, contrary to the above argument a significant number of novae have been observed to form dust in their ejecta after the outbursts. Generally, the formation of dust is inferred from a dip observed in their optical light curve due to the obscuration of light and rise in their infrared emission, e.g., DQ Her (McLaughlin 1935), FH Ser (Geisel et al., 1970), NQ Vul (Ney \& Hatfield 1978), Nova Serpentis 1978 (Gehrz et al., 1988), QV Vul (Gehrz et al., 1992), V842 Cen (Andrea, Drechsel \& Starrfield 1994), V1419 Aql (Lynch et al., 1995), V705 Cas (Evans et al., 1996), V1280 Sco (Das et al., 2008), V1065 Cen (Helton et al., 2010), V476 Scuti (Das et al., 2013), V2676 Oph (Kawakita et al., 2017), V339 Del (Shore et al. 2018; Skopal 2019), V5668 Sgr (Shore et al. 2018), etc. Dust formation is also evidenced from the asymmetries observed in the line profiles. For example, in the study of the dusty nova V5668 Sgr, Shore et al. (2018) observed asymmetry in the line profiles with suppression of red wing during the beginning of the dust formation epoch, just prior to the onset of deep minimum in the optical and ultraviolet (UV) light curves. In the case of V705 Cas, it was observed that the UV (1200 - 3000 \AA) flux was decreased by more than an order of magnitude but the spectrum remained unchanged in line ratios, while a rapid rise in IR emission was observed (Shore et al. 1994). The deficit in the UV flux was balanced by the IR when the grain optics were taken into account. The authors showed that the dust formed in the outer region of the ejecta at a large distance $\sim 10^{14} - 10^{15}$ cm from central ionizing WD and the absorbed flux was redistributed.

Novae dust forms within a short time duration of $\sim$ 30-100 days after an outburst (Evans 1990), in comparison to the interstellar dust which typically takes a few thousand years to form. Observations show that the composition of novae dust includes carbon, silicates, SiC, and hydrocarbons, and sometimes, a combination of these (Gehrz et al., 1998; Evans \& Gehrz 2012). It is found that novae grains can grow as large as $\sim$ 0.2 to 8$\mu$m (Evans et al., 2017), larger than the grains in the interstellar medium which have radius $\leq 0.2 \mu$m  (Gehrz 1999; Evans et al., 1997; Shore et al., 1994). However, dust formation in the hostile environment in novae ejecta has been an open question for many decades. Several attempts have been made to understand the physical and chemical conditions required to form dust in novae ejecta, and its relation with the observable parameters. Each of the studies depends on the specific scenario, such as the relation between speed class and dust formation (Gallagher 1977; Bode \& Evans 1982), the effect of enrichment of CNO elements on dust formation (Gehrz 1988, and references therein), the association of dust with CO emission (Rudy et al. (2003), the time it takes for the brightness of a  nova to decline by two magnitudes, t$_2$ (Shafter et al. 2011). Shore \& Gehrz (2004) proposed the ionization-mediated kinetic agglomeration of atoms onto molecules and small grains through induced dipole interactions. The role of shock on the chemistry of dust formation in novae has been discussed by Derdzinski et al. (2017). Recently, Shore et al. (2018) studied the development of spectral line profiles with the evolution and distribution of dust grains in biconical ejecta and argued that the dust does not destroy even after the ejecta becomes transparent to the UV and harsh X-ray radiations from the central ionizing source and thereby the dust grains are likely to survive until mixed with ISM. However, due to the inherent complexity of the physical and chemical composition of novae ejecta and the multi-step process of dust grain formation, such attempts which relied only on one or two observed quantities, could only provide partial success in explaining the dust formation. Thus, a more fundamental approach is required where multiple physical and chemical parameters of the dust forming novae ejecta are studied in detail, to estimate favorable conditions for the formation of dust.

In the present work, we model observed pre-dust and post-dust spectra of the dust forming nova V1280 Scorpii (V1280 Sco), as a paradigmatic case, to study the dust formation phenomenon in a nova. Our primary aim is to understand how the crucial physical parameters associated with the system evolve with the formation of dust. Keeping this in mind, we construct a simple phenomenological model assuming a spherical geometry using \textsc{cloudy}, v.17.02 (Ferland et al., 2013, 2017) with the preliminary model parameters adopted from the previously published studies. To include the effect of dust, we add dust grains in the outer shell of the ejecta in the post-dust phase. Such detailed photoionization modeling of dust forming nova has not been done earlier. Our paper is organized in the following way: a brief introduction on V1280 Sco and observational findings are summarized in Section \ref{v1280sco}. Details of observations are described in Section \ref{dataset}. In Section \ref{cloudy}, we briefly describe the Photoionization code \textsc{cloudy} and  the modeling procedure. The results are discussed in section \ref{results} followed by the summary in Section \ref{summary}.

\section{V1280 Sco (Nova Scorpii 2007)}\label{v1280sco}
\begin{figure}
\centering
\includegraphics[scale=0.5]{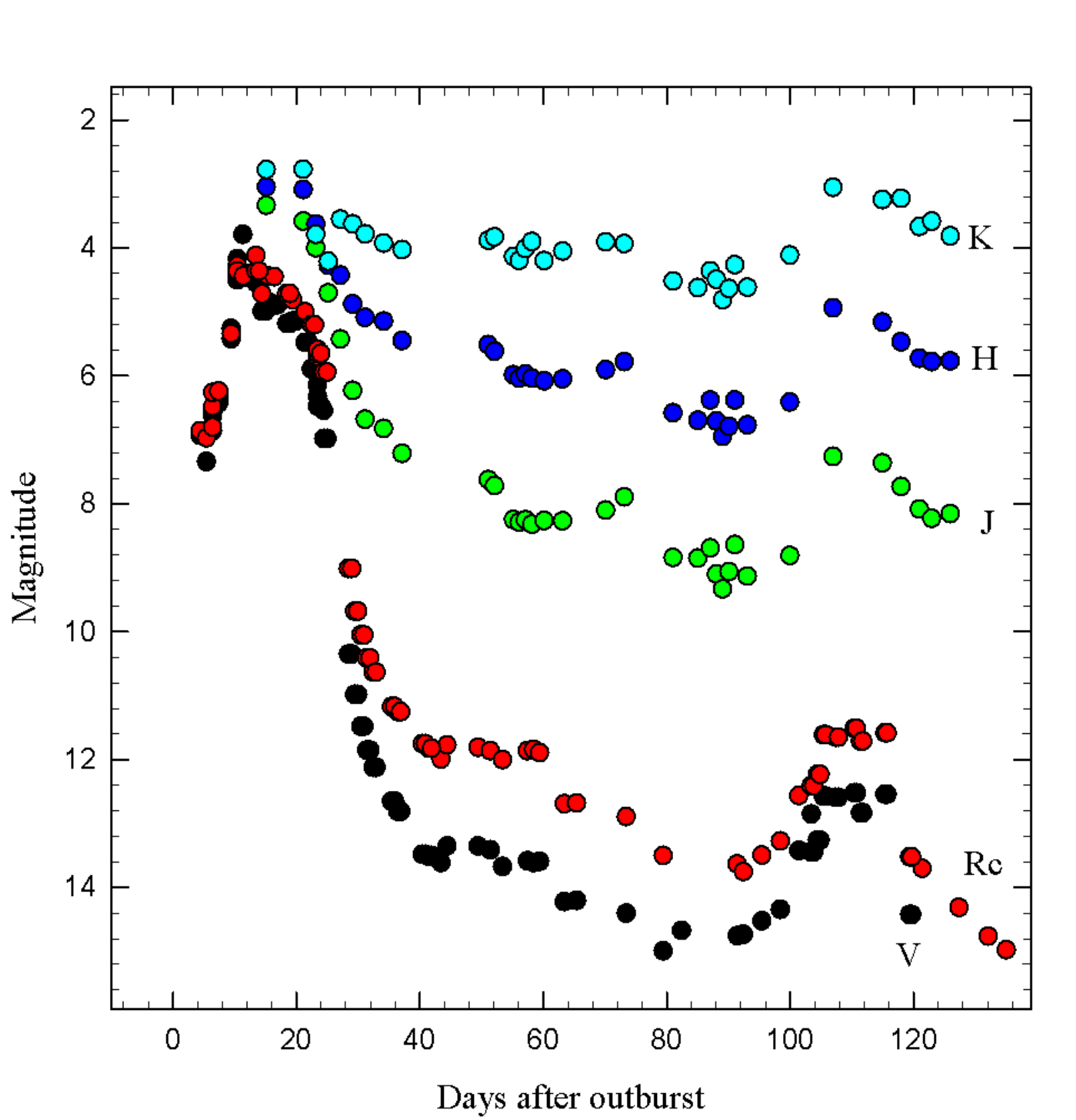}
\caption{The light curve of V1280 Sco in multiple wavelength bands for comparision (V band - black circles, R$_c$ band - red circles, J band - green circles, H band - blue circles, and K band - cyan circles, from Das et al., 2008). The data obtained from Mt. Abu observations, IAU circulars, VSNET alerts, VSOLJ and AFOEV databases ((Variable Stars Network, Japan; Variable Star Observers League in Japan; Association Francaise des Observateurs d'Etoiles Variables, France)}
\label{fig1}
\end{figure}

V1280 Scorpii (V1280 Sco) is a classical nova that was discovered in outburst on 2007 February 4.8 (JD =2 454 136.85). In-depth studies of the nova in optical and near-Infrared (NIR) regions have been done by Naito et al. (2012), and Das et al. (2008), respectively. The nova reached maximum brightness, V = 3.8, on Feb. 16, about 12 days after the outburst. V1280 Sco is an extremely slow nova with t$_2$ $\sim$ 21 days (Das et al., 2008), the nebular phase was entered about $\sim$ 1600 days after an outburst (Naito et al. 2012). Chesneau et al. (2008) studied the early phases of V1280 Sco using interferometric data observed at VLTI/AMBER and MIDI between days 23 and 145 and detected an expanding dust shell moving at $0.35 \pm 0.03$  mas day$^{-1}$. As the interferometric observations provided little information on the shape of the source, Chesneau et al. (2008) considered a spherical geometry of the ejecta in their studies. They determined typical velocities $\sim$ 500 km s$^{-1}$ from the typical P-cygni profile and estimated a distance of 1.6 $\pm$ 0.4 kpc. Later, high angular resolution infrared observations of V1280 Sco with the NACO/VLT and VISIR/VLT in 2009, about two years after the outburst, revealed bipolar-shaped nebula around V1280 Sco, with dust present mostly in the lobes of the nebula. The key feature is that the nova formed thick dust $\sim$ 24 days after the outburst, which was evidenced from the deep minimum in visual light curve occurred due to the obscuration by dust shell, and rise in near-Infrared JHK fluxes (see fig \ref{fig1}).

The pre-dust phase optical spectrum (see Figure~\ref{fig:ep02}) is mostly dominated by prominent hydrogen Balmer lines such as H$_{\alpha}$, H$_{\beta}$, H$_{\gamma}$, and H$_{\delta}$ along with the Fe II lines. In addition, the Na I D1, D2 lines and forbidden lines of [O I] at 5577 \AA,~6300 \AA, and 6364 \AA,~respectively, are also observed in the optical spectrum on 28 Feb 2007 at very low strength. The [O I] emission in optical spectra indicates the presence of a neutral zone (Anupama \& Kamath 2012). These neutral lines of [O I] and the Na I are observed throughout the Fe-curtain peak opacity and lifting stages (Shore 2019). No line of He I is seen in the early optical spectra of the nova. Two features of He I (5876 and 6678 \AA) were seen in the optical spectrum observed on 09 September 2008, $\sim$ 571 days after the outburst (Naito et al. 2012). The near-Infrared JHK spectra of V1280 Sco are presented in Figures~\ref{fig:ep01},~\ref{fig:ep02},~\ref{fig:ep03}, and~\ref{fig:ep04} in this paper. The J-band spectra of the pre-dust phase are dominated by the prominent lines of Pa$\gamma$ 1.0938$\mu$m, He I 1.0830$\mu$m, blend of O I 1.1287$\mu$m and C I 1.11330 $\mu$m in comparable amount, a complex blend of several C I and N I lines between 1.2 and 1.275 $\mu$m at very low strength, Pa $\beta$ 1.2818$\mu$m, O I 1.3164$\mu$m. In the H band, the recombination lines of the Brackett series, especially Br 10-12 are prominent though they are heavily blended with C I lines. Apart from the hydrogen lines, the strong C I emission at 1.6005$\mu$m, 1.6890$\mu$m, 1.7234 - 1.7275$\mu$m, and in the region between 1.7449 to 1.7814$\mu$m are seen. In K-band, Br $\gamma$ at 2.1655 $\mu$m is the strongest line; other noticeable lines are He I 2.0585 $\mu$m, and several blended C I features at 2.1156 - 2.1295$\mu$m, 2.2156 - 2.2167$\mu$m, 2,2906$\mu$m, and 2.3130$\mu$m. Besides, several lines of Mg I (1.1828 $\mu$m, 1.5040$\mu$m, and 1.5749$\mu$m, and 1.7109$\mu$m) and Na I (1.1381, 1.1404, 2.2056, 2.2084 $\mu$m,) are also present in the JHK spectra -  which indicates the presence of a low-temperature region in the nova ejecta (Das et al. 2008). The prominent lines of C I indicate that the novae belong to CO type novae (Das et al., 2008; Banerjee \& Ashok, 2012). Detailed lists of the observed spectral lines, in optical and NIR bands, are presented in Naito et al. (2012) and Das et al. (2008), respectively.

Observations show the evidence of the formation of simple diatomic molecules in the nova ejecta (Evans \& Rawlings 2008). For example, CN was the first molecule observed in the optical spectrum of DQ Her nova (Wilson \& Merrill, 1935). Optical contributions by C$_2$ and CN are also observed in a dust-forming nova V2676 Oph (Kawakita et al. 2017). H$_2$ (2.122 $\mu$m) was also detected in the DQ Her at much later epochs (Evans 1991). C$_2$ and H$_2$ could be an essential intermediate in the formation of carbon dust in novae (Pontefract \& Rawlings 2004). The first overtone of CO has also been seen in the near-infrared spectrum of a handful of novae, e.g., NQ Vul (Ferland et al. 1979), V705 Cas (Evans et al. 1996), V2274 Cyg (Rudy et al. 2003), V2615 Oph (Das et al. 2009), V496 Sct (Raj et al. 2012), V5668 Sgr (Banerjee et al. 2016); V6567 Sgr (Woodward et al. 2020), etc. The presence of CO in novae indicates the presence of a high-density, low temperature, and metal-rich environment, which could be also a pre-requisite condition for the formation of dust (Rudy et al. 2003; Das et al. 2008). However, all dust-forming novae have not shown the CO emission in their spectra. In the case of V1280 Sco also, no CO emission was detected in the observed spectra (Das et al. 2008; Banerjee \& Ashok 2012).  A more detailed discussion on the chemistry of dust and molecules in novae ejecta is provided in Evans \& Rawlings (2008).

\section{Data Set}\label{dataset}
In the present analysis, we model the optical spectra (from Naito et al., 2012) and near-infrared (NIR) $JHK$ spectra (from Das et al., 2008). Two epochs of pre-dust phase: 24 February 2007 (hereafter, \enquote{epoch 1}) and 27-28 February 2007 (hereafter, \enquote{epoch 2}), and two epochs of post-dust phase, 03 March 2007 (hereafter, \enquote{epoch 3}) and 07 March 2007 (hereafter, \enquote{epoch 4}) are used for modeling. Optical medium-dispersion observation on 28 Feb 2007 was carried out using three instruments, the MALLS (Medium And Low dispersion Long slit Spectrograph) mounted on the 2.0-m NAYUTA telescope at Nishi-Harima Astronomical Observatory (NHAO). The spectral resolution at H$\alpha$ was R $\sim$ 1200 (Naito et al., 2012). NIR observations of V1280 Sco were obtained at the 1.2 m telescope in Mt. Abu telescope at dispersions of $\sim9.75 \AA$  pixel$^{-1}$ using 256 $\times$ 256 HgCdTe Near-Infrared Camera Multiobject Spectrograph 3 (NICMOS3) array (Das et al., 2008).

\section{Modelling procedures}\label{cloudy}
We use the photoionization code \textsc{cloudy}\footnote{\url{https://www.nublado.org}}, v.17.02 (Ferland et al. 2017) to model the observed spectra of V1280 Sco. This code is based on detailed microphysics to simulate the physical conditions of non-equilibrium gas clouds exposed to an external radiation field. It solves the thermal, statistical, and chemical equilibrium equations self-consistently. Currently, it uses 625 species including atoms, ions, and molecules and five distinct databases: H-like and He-like iso-electronic sequences (Porter et al., 2012), Stout (Lykins et al., 2015), \textsc{chianti} (Landi et al., 2012), \textsc{lamda} (Schoier et al., 2005) and the $H{_2}$ molecule (Shaw et al., 2005) to model spectral lines. All the known important ionization processes, e.g., photo, Auger, collisional and charge transfer and recombination process viz. radiative, dielectronic, three-body recombination and charge transfer are included self-consistently. \textsc{cloudy} predicts both the intensities and column densities of a very large number ($\sim 10^{4}$) of spectral lines covering the whole electromagnetic range, from non-LTE (Local Thermodynamic Equilibrium), illuminated gas clouds by solving the equations of thermal and statistical equilibrium for a given set of input parameters. More details about \textsc{cloudy} could be found in Ferland et al. (2013), Shaw et al. (2020), and references therein.

We consider a central ionizing source surrounded by a spherically symmetric, expanding gaseous ejecta whose dimensions are determined by the inner (R$_{in}$) and outer (R$_{out}$) radii (cm). The central ionizing radiation is assumed to have a blackbody shape with temperature T$_{\text{BB}}$ (K) and luminosity L (ergs$^{-1}$). We assume the surrounding gas to be clumpy. Here clumpiness describes the fraction of volume that contains gas. The clumpiness is set by filling factor, $f(\text{r})$. In the present work, we use a simplistic approximation for clumps using filling factor in our model, and this is the only available option in \textsc{cloudy} for 1D modeling as of now. Its value varies with the radius as given by the following relation,
\begin{equation}
f(\text{r})=f(\text{r}_{in})(\text{r/r}_{in})^{\beta}
\end{equation}
where $\beta$ is the exponent of the power-law. The filling factor in novae ejecta has small values, which lie in the range from 0.01 to 0.1 (Shore 2008). During the early stage spectra of nova ejecta, the value remains $\sim$ 0.1  which decreases with time (Ederoclite et al. 2006).  Since we model the early stages of the V1280 Sco, we consider the value of filling factor to be 0.1, and $\beta = 0.0$ for our model calculations, which is also similar to the previous studies (e.g., Della Valle et al. 2002; Vanlandingham et al. 2005; Helton et al. 2010; Mondal et al. 2018; Pavana et al 2019 etc). This is similar to the value of $f$ estimated as $\sim$ 0.1, using the integrated H$\beta$ flux of the dust forming nova V339 Del in the late stages (Shore et al. 2016).
In addition, the fraction of the 4$\pi$ (steradian) sr of the gas irradiated by the central source is parameterized by covering factor whose value lies in the range from 0 to 1. The density of the ejecta is set by the total hydrogen number density, $n(\text{H})$ [cm$^{-3}$] given by,
\begin{equation}
n(\text{H}) = n(\text{H}^{0}) + n(\text{H}^{+}) + 2n(\text{H}_{2}) + \sum_{\text{other}} n(\text{H}_{\text{other}})~ \text{cm}^{-3},
\end{equation}
where, $n(\text{H}^{0})$, $n(\text{H}^{+})$, $ 2n(\text{H}_{2})$, and $n(\text{H}_{\text{other}})$ represent hydrogen in neutral, ionized, molecular, and all other hydrogen bearing molecules, respectively. The elemental abundances of the ejecta, relative to hydrogen, are set by the \textit{abundance} parameter. Furthermore, we use a radius dependent power-law density distribution, $\rho \propto \text{r}^{\alpha}$ with power-law index $\alpha$ given by Bath \& Shaviv (1976),
\begin{equation}\label{eq:one}
n(\text{r})=n(\text{r}_{\text{in}})(\text{r/r}_{\text{in}})^{\alpha} \text{cm}^{-3}
\end{equation}
where, $n(\text{r})$, and $n(\text{r}_{\text{in}})$ are the density of the ejecta at the distance r and the
inner radius (r$_{\text{in}}$), respectively.
For a steady-state condition, density $\rho$ is given by,
\begin{equation}
\rho = \frac{\dot{\text{M}}}{4 \pi \text{r}^2 \text{v}}
\end{equation}
where, $\dot{\text{M}}$, and $\text{v}$ are the mass outflow rate and outflow velocity, respectively.

Spherically symmetric, line blanketing non-LTE expanding atmospheric models of novae have been developed over the past few decades (see Hauschildt et al. 1992; 1994; 1996). In the very early stages of the nova ejecta, a very sharp density profile  ($\alpha \sim 10$) was used in the models (Hauschildt et al. 1994; Short et al. 2001; Hauschildt 2008). As the ejecta expands adiabatically, the temperature decreases from $\sim$ 15,000 K to $<$ 10,000 K. This expansion also reduces the optical depth, making the deeper layers of the ejecta visible. During this stage, the density profile of ejecta becomes shallower ($\alpha \sim 3$). This density profile implies that the nova atmosphere has a very large geometrical extension and thereby a very large range of temperatures and ionization states within the continuum and line-forming regions (Hauschildt 2008; Chomiuk, Metzger \& Shen 2020). Assuming that the kinematics of the novae ejecta could be consistently explained by clumpy gas expelled during a single ballistic or free expansion (Mason et al. 2018, Shore et al. 2018) which follows linear velocity law (v $\propto$ r), we adopted the $\alpha = 3$ for our model calculations, a typical value which was used in the photoionization modeling of other novae (e.g. Schwarz 2002; Vanlandingham et al., 2005; Helton et al., 2010; Das \& Mondal, 2015; Mondal et al., 2018; Pavana et al. 2019, etc).

We used n(H), T$_{BB}$ and L of the central ionizing source, covering factor for both clump and diffuse parts (see section 4.1 for discussions), abundances of the elements as the input free parameters. We considered those elements only whose emission features are present in the observed spectra and kept others at the solar values (Grevesse et al. 2010). We spanned a large sample space for all input parameters varying their values in smaller steps. Initially, we varied T$_{BB}$ (in the range from 10$^{4.0}$ to 10$^{5.0}$ K), Luminosity (in the range from 10$^{35.5}$ to 10$^{37.5}$ ergs$^{-1}$) and n(H) (in the range from 10$^{9.0}$ to 10$^{14.0}$ cm$^{-3}$),  keeping the elemental abundances at the solar values, to obtain a best visual match of the continuum and prominent H lines of the observed spectra. The upper and lower limits of the above parameters were chosen based on the results obtained in the previous studies. After obtaining a reasonably good match we focused on the C I and O I lines which are the most prominent lines other than Hydrogen, and the weak features of nitrogen, sodium, magnesium, and iron in the spectra, and we varied C, N, O, Na, Mg, and Fe abundances, simultaneously. While doing so we also allowed moderate variations of the T$_{BB}$, L, and n(H). We followed the same procedure to generate model spectra for different sets of parameters for other epochs. For the post-dust phase, we had to carefully match the rise in continuum in addition to the emission lines by adding dust grains of different sizes in the model. The entire process of running each case (for both clump and diffuse parts) has been done manually. To generate all prominent C I lines we had to include all available energy levels of \textsc{stout} database. In this way, we ran more than twenty five thousands cases and generated spectrum for each case. Initially, we matched the observed and modeled spectra by visual inspection. We chose the reasonably good fits and discarded those model spectra which deviated significantly from the observed spectra. We then calculated the $\chi^2$ and $\chi_{red}^{^2}$ using the fluxes of modeled and observed spectra with $\ge$ 30 emissions lines, to get the best-fit for each epoch.

The $\chi^2$, goodness of fit of a model to the observed spectrum is determined by the following relation,
\begin{equation}
\chi^2 =\sum_{i=1}^n\frac{(M_i-O_i)^2}{\sigma_i^2}
\end{equation}
where $n$ is the number of emission lines used in the model, $M_i$ is the modeled line flux ratios, $O_i$ is the observed line flux ratios, and $\sigma_i$ is the error in the observed line flux ratios. The reduced $\chi^{2}$ is given by,
\begin{equation}
\chi^{2}_{\text{red}} = \frac{\chi^2}{\nu}
\end{equation}
where, $\nu$ is the number of degree of freedom (DOF), given by the difference between the number of observed emission lines ($n$) and the number of free parameters ($n_p$), i.e., $\nu$ = $n - n_{p}$. For the acceptable fitting, the value of $\chi^2$ $\sim$ $\nu$ and the $\chi^2_{red}$ should be low, typically, in between 1 and 2. The value of $\sigma$ generally lies in the range of 10\% to 30\% (Helton et al. 2010) depending on the strength of spectral line relative to the continuum and the possibility of blending with the other lines. In the present analysis, we use the integrated line flux of V1280 Sco as the line profiles are unknown and are heavily blended. Thus, we consider $\sigma$ = 30\% for the present study. The best-fit model parameter values with the uncertainties are presented in Table~\ref{tab:fit}. To determine uncertainties associated with the parameters, we vary one parameter at a time keeping other parameters fixed at their best-fit model values until the $\chi_{red}^2$ of the model increased to 2. This approach gives approximately 3$\sigma$ uncertainty for the free parameters (Schwarz et al. 2001; 2007). The variation of reduced Chi-square with different free parameters are presented in Figure~\ref{fig:ep07}.

\begin{figure}[ht!]
\centering
\includegraphics[scale=0.25]{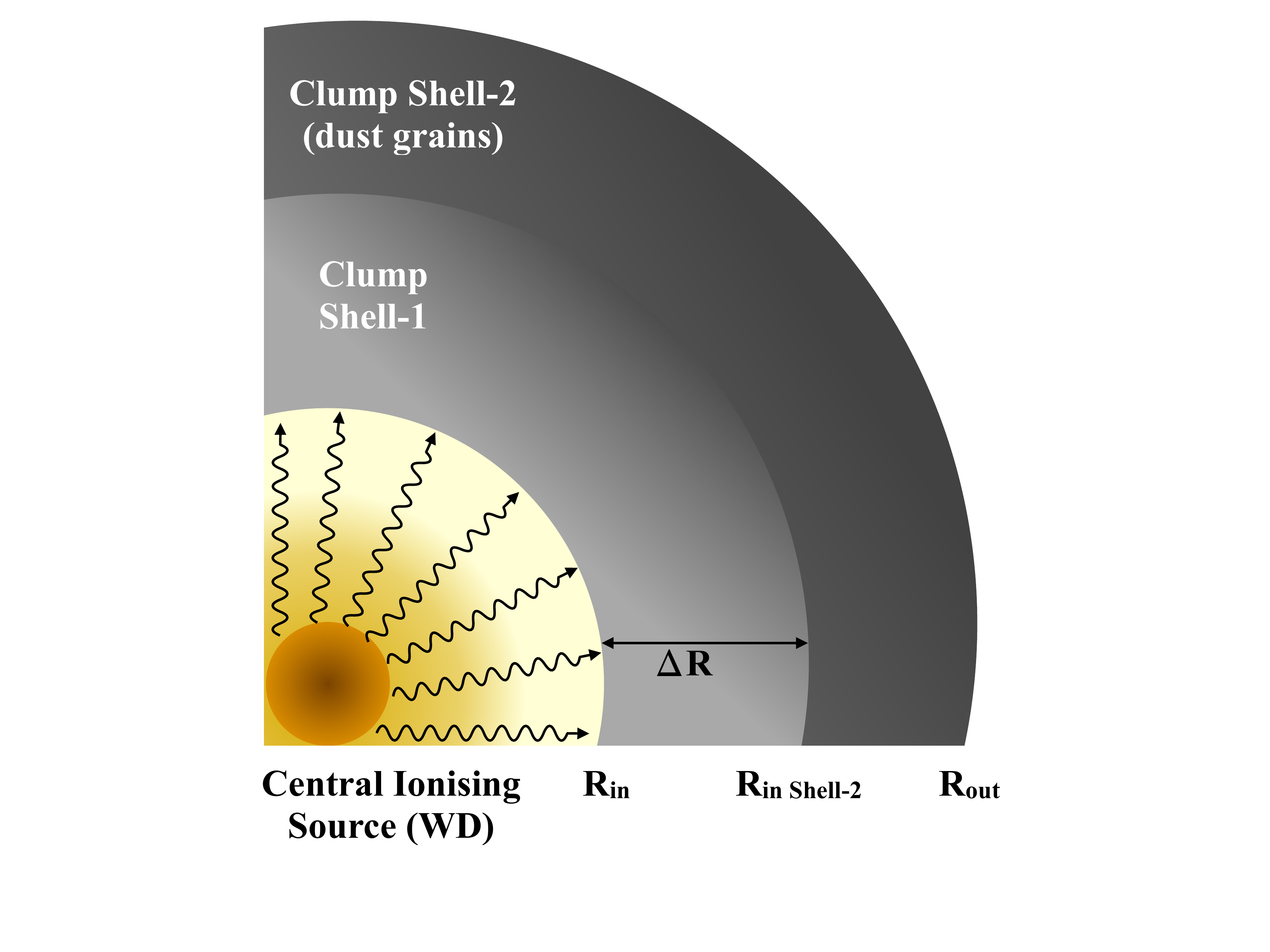}
\caption{A Schematic diagram for post-dust phase model. See section~\ref{result:dust} for detailed discussions}
\label{cartoon}
\end{figure}

The spectra of V1280 Sco show evidence of dust formation at the later stage of evolution. The crucial condition for dust formation is to have dust temperature below its sublimation temperature. The sublimation temperature for graphite and silicate is about 1750 K and 1400 K, respectively. In \textsc{cloudy}, the gas-phase abundances are not automatically changed where the grain abundances are changed. Hence, we consider two separate cases: pre-dust and post-dust formation phases as discussed below. The emergent continuum from the pre-dust formation phase is used as the incident continuum for the post-dust formation phase. Figure~\ref{cartoon} depicts a schematic diagram of our two-phase model.

\subsection{Pre-dust formation phase}\label{Pre-dust}
We model the pre-dust formation phase (epochs 1 and 2) first. In a previous study, Das et al. (2008) used a simple LTE model to generate the NIR spectra; with the primary aim of identifying the spectral lines only. The purpose of the present work is to estimate the physical parameters associated with the nova on different epochs. Because of the complex and inhomogeneous nature of nova ejecta, a single component model can not generate all the lines properly. Hence, we assume that the nova ejecta is composed of at least two different density regions - one diffuse region with the low hydrogen density to generate the emissions lines of high ionization potential and one clump region with relatively high hydrogen density to generate the emission lines of lower ionization potential. It is assumed that the ejecta has two density regions with different covering factors, the sum of which is less than or equal to 1. The value of filling factor and $\beta$ are set to 0.10 and 0, respectively. The chemical abundances of all elements are held fixed at their solar values (Grevesse et al. 2010), relative to hydrogen, except for He, C, N, O, Na, Mg, and Fe, as lines of these elements are present in the observed spectra. We assume that the elemental abundances are the same in both regions. Similar two-component modeling has been used successfully in the studies of other novae systems, e.g., Nova LMC 1991 (Schwarz et al. 2001) QV Vul (Schwarz 2002), V382 Vel (Shore et al. 2003), V1974 Cyg (Vanlandingham et al., 2005), V838 Her \& V4160 Sgr (Schwarz et al., 2007), T Pyx (Pavana et al., 2019), etc. To check the role of each parameter on the spectra we also run the code for individual parameters (e.g. temperature and luminosity of central ionizing source, ejecta size, and density, elemental abundances of ejecta, etc.) keeping others fixed. We observed that our model is susceptible to various input parameters, a small change in any of these parameters causes a noticeable change in the relative line intensities. Thus these spectral features helped in estimating the values of these parameters; we discussed these in the results section in detail.

In the previous studies, Naito et al., (2012) estimated the expansion velocity of the dust shell to be $350 \pm 160$ km s$^{-1}$ from the spectral line analysis. From extensive interferometric observations of the early phase of the outburst, Chesneau et al. (2008) determined a more precise value of the expansion velocity as $500 \pm 100$ km s$^{-1}$ and interpreted a spherical shell geometry. Considering these findings, in the present study, we adopt a spherical geometry and expansion velocity of the ejecta as $500$ km s$^{-1}$ as the spectra used in this work were observed in the early phase, in between 20 and 31 days after the outburst. We calculate the initial radius of the ejecta; R$_{in}$ = $7.94 \times 10^{13}$ cm, $8.91 \times 10^{13}$ cm, $1.04 \times 10^{14}$ cm, and $1.23 \times 10^{14}$ cm, for epoch 1, epoch 2, epoch 3, and epoch 4, respectively. The nova atmospheres are optically thick and fast-expanding shells with a very slow decrease of density profiles with increasing radius. This leads to very large geometrically extended atmospheres having large temperature differences between the inner and outer parts of the line-forming regions. Typically, the relative geometrical extension, R$_{out}$/R$_{in}$ of a nova atmosphere is $\sim$ 100 - 1000, which is much larger than the geometrical extension of hydrostatic stellar atmospheres (see e.g., Hauschildt et al. 1992, 1995). Thus we made the resulting curvature of our model substantially large and also observed that the model is mostly insensitive to the choice of outer radius. The ejecta might be radiation-bounded which could be inferred by the presence of [O I] lines present in the optical spectra which originates in the region where neutral oxygen is present (Ferland 1977). The cooler outer layers of the ejecta allow dust to form at a larger distance from the central ionizing WD (Shore et al. 1994). However, in our model analysis, instead of outer radius we have used a certain value of ejecta temperature as a stopping criterion for the \textsc{cloudy} calculations. We discuss this in detail in section~\ref{result:dust}.

\subsection{Post-dust formation phase}\label{post-dust}
In order to model the post-dust phase (epochs 3 and 4) spectra, we added grains in our two-component model. Generally, the astrophysical dust grains in the ISM are spherical and follow a classical MRN (Mathis, Rumpl \& Nordsieck, 1977) power law distribution given by;
\begin{equation}
n(a) \propto a^{q} ~,~ a_{min} \leq a \leq a_{max}
\end{equation}
where, $a_{max} = 250~nm$, $a_{min} = 5~nm$, and the power of index $q$ = -3.5. We assume a similar distribution form for novae dusts with different values of $a_{min}, a_{max}$, and $q$ (see section~\ref{result:dust} for detailed discussions). Novae grains are usually larger in size ($\sim 0.5 \mu$m) than the ISM grains ($< 0.2\mu$m), and they decrease to the dimensions of ISM grains rapidly by sputtering and/or evaporation (Evans et al. 1997; Sakon et al. 2016). 
We use the net transmitted continuum of the pre-dust phase as the incident continuum for the post-dust phase. The inner radius, in this case, is the sum of the initial radius of the ejecta ($1.04 \times 10^{14}$ cm) and the thickness of the pre-dust ejecta. We use equation \ref{eq:one} to estimate the value of hydrogen density at the inner radius of the post-dust phase. The outer edge of the post-dust ejecta extends to the radius where the gas temperature falls to 10 K. The resulting spectrum is obtained by adding the spectrum from clumps and that from the diffuse part, after multiplying with their corresponding covering factors. To the best of our knowledge, this seems to be the first attempt to model spectra of any dusty nova using \textsc{cloudy}, incorporating dust grains in the ejecta shells.

\begin{figure}[ht!]
\gridline{\fig{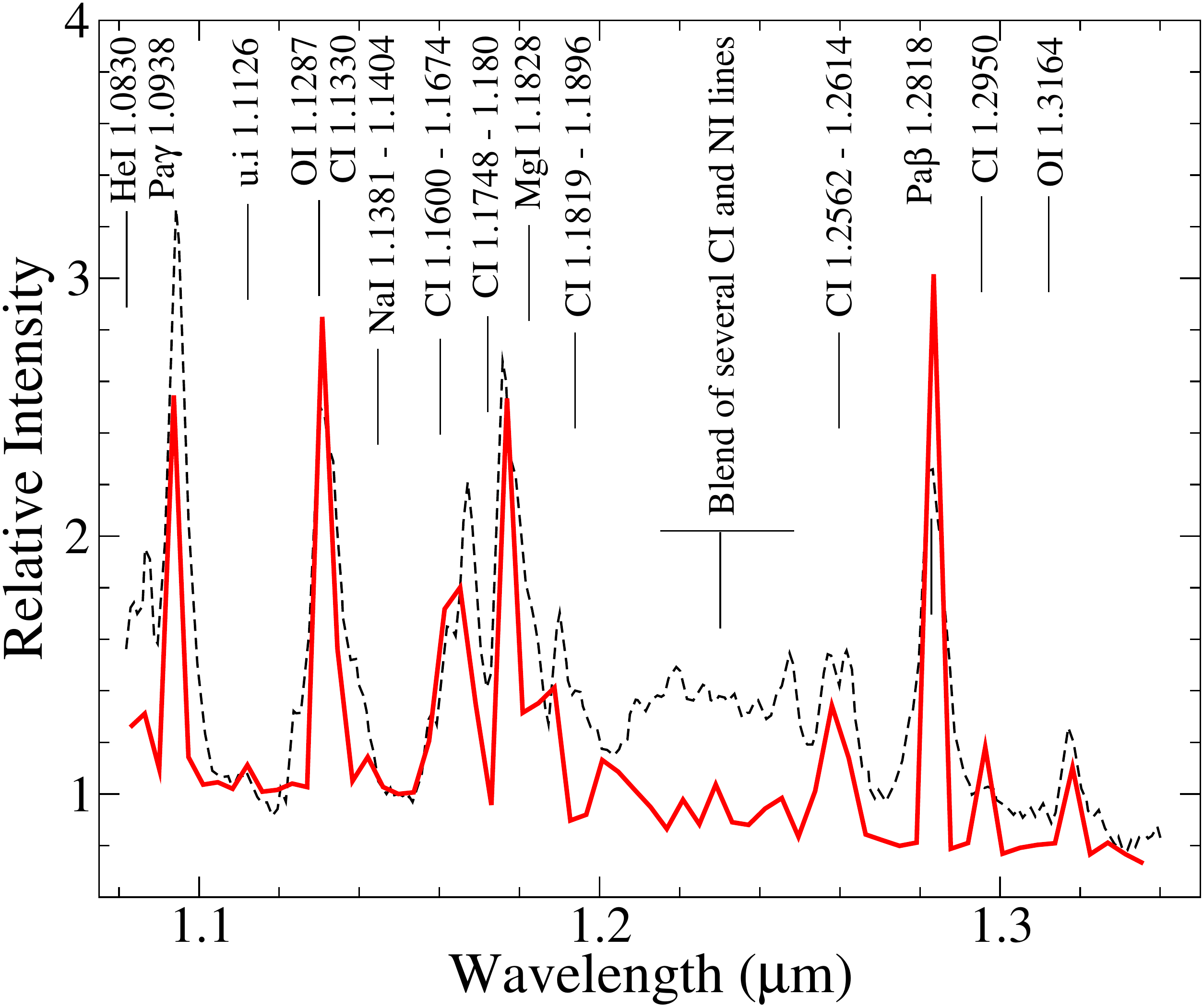}{0.4\textwidth}{(a)}}
\gridline{\fig{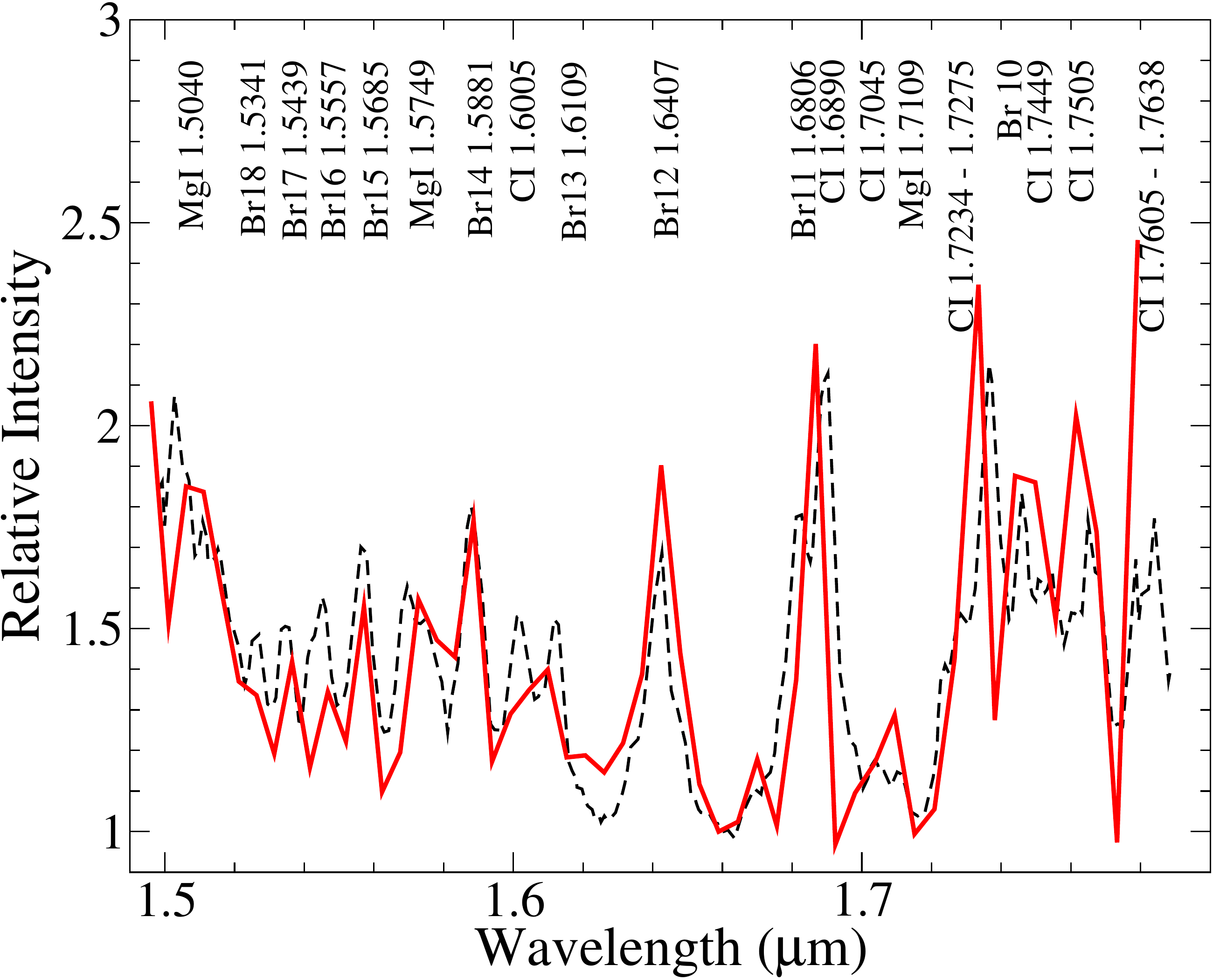}{0.4\textwidth}{(b)}}
\gridline{\fig{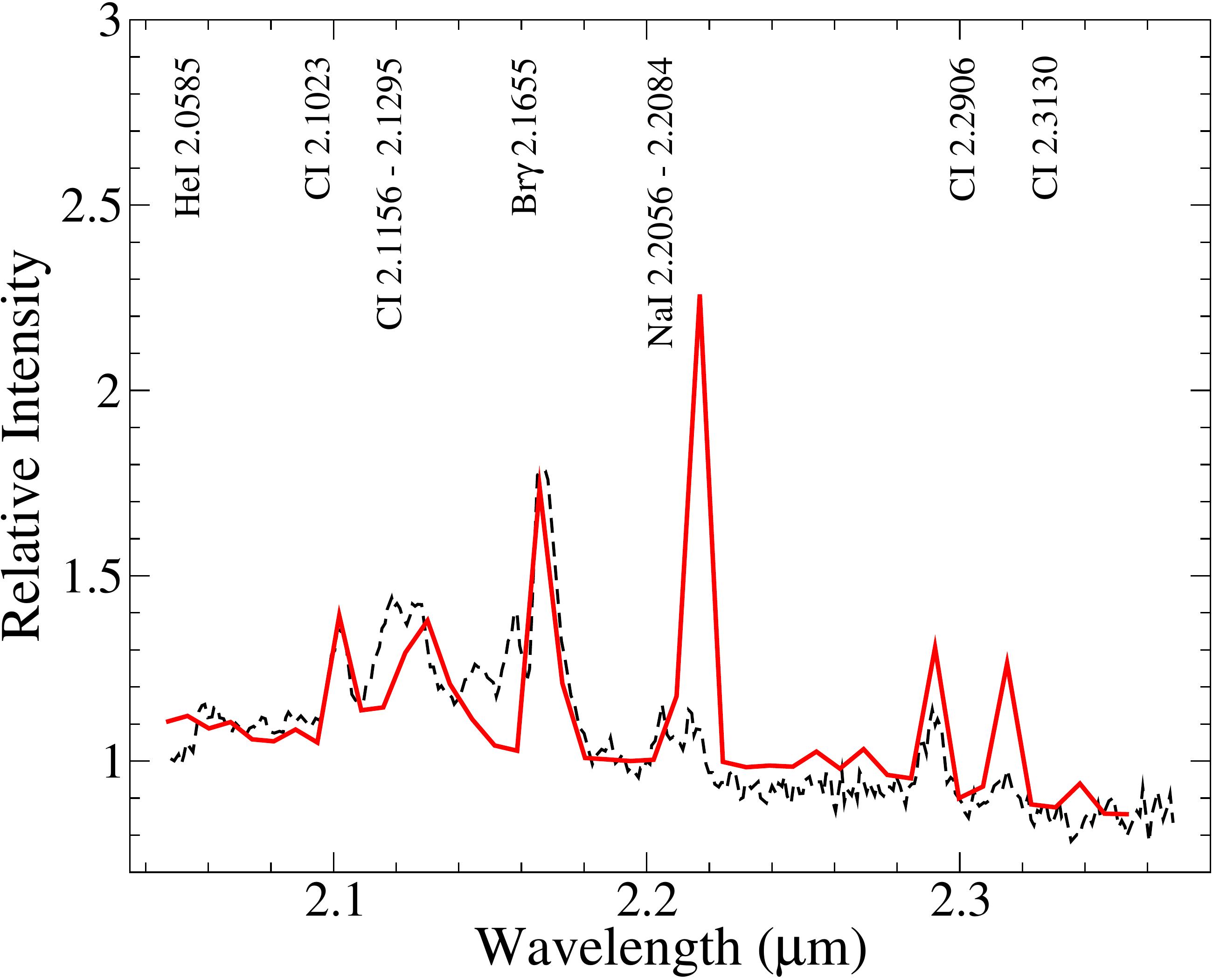}{0.4\textwidth}{(c)}}
\caption{Observed (black dashed solid) and model generated spectra (red solid line) of epoch 1 (pre-dust phase). (a) J-band. (b) H-band. (c) K-band. For more details see section~\ref{results}}
\label{fig:ep01}
\end{figure}

\begin{figure*}[ht!]
\gridline{\fig{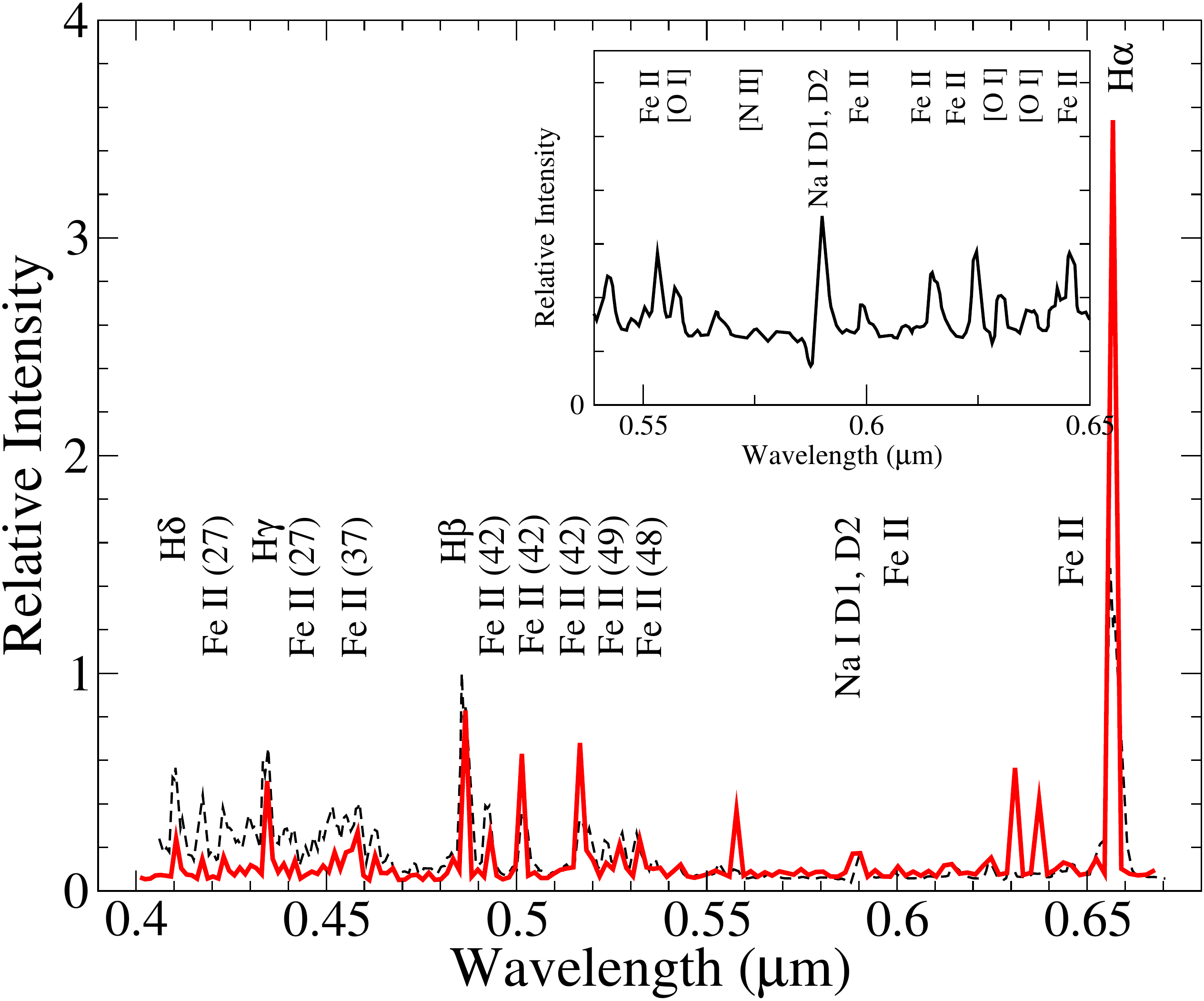}{0.4\textwidth}{(a)}
         \fig{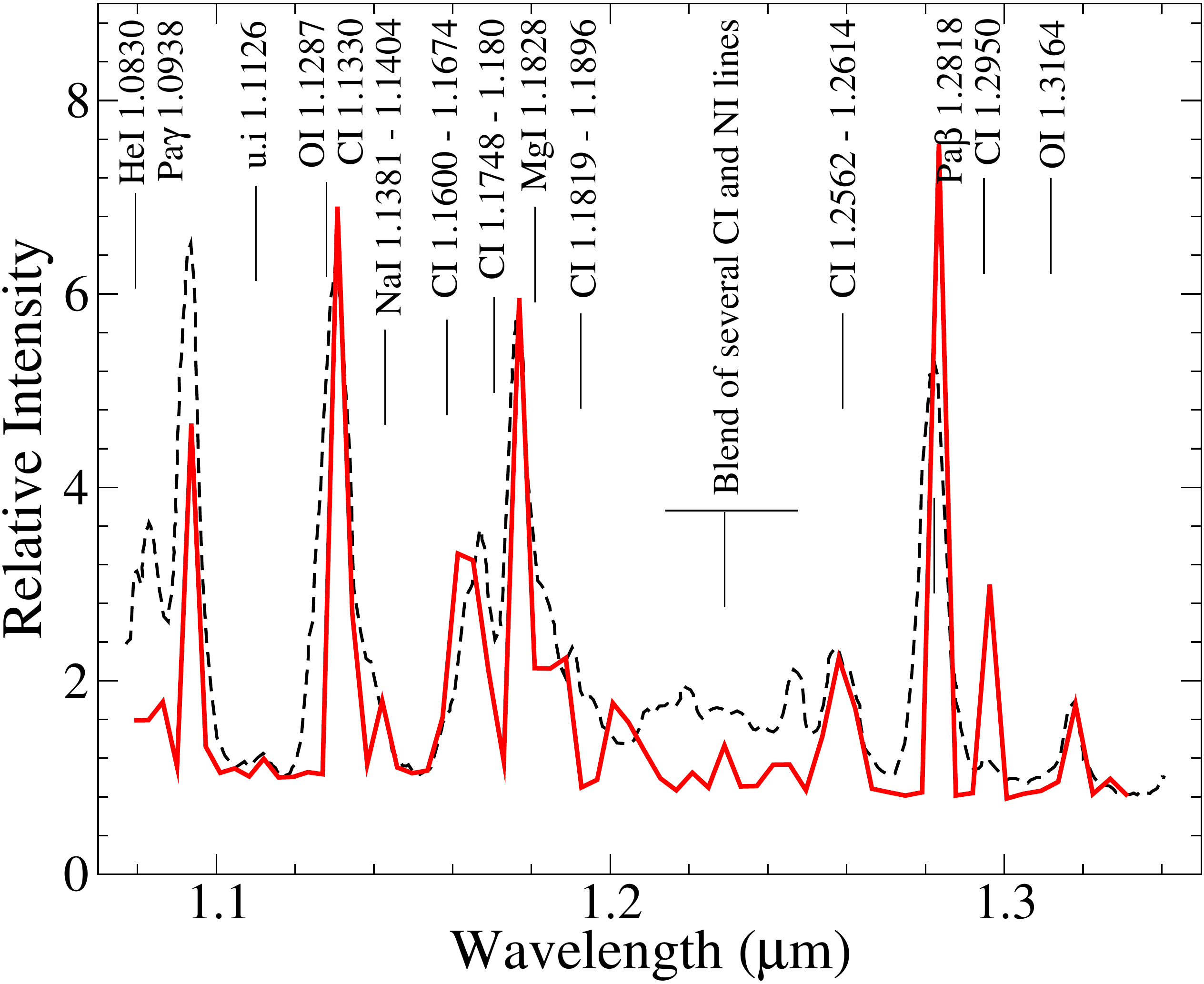}{0.4\textwidth}{(b)}}
\gridline{\fig{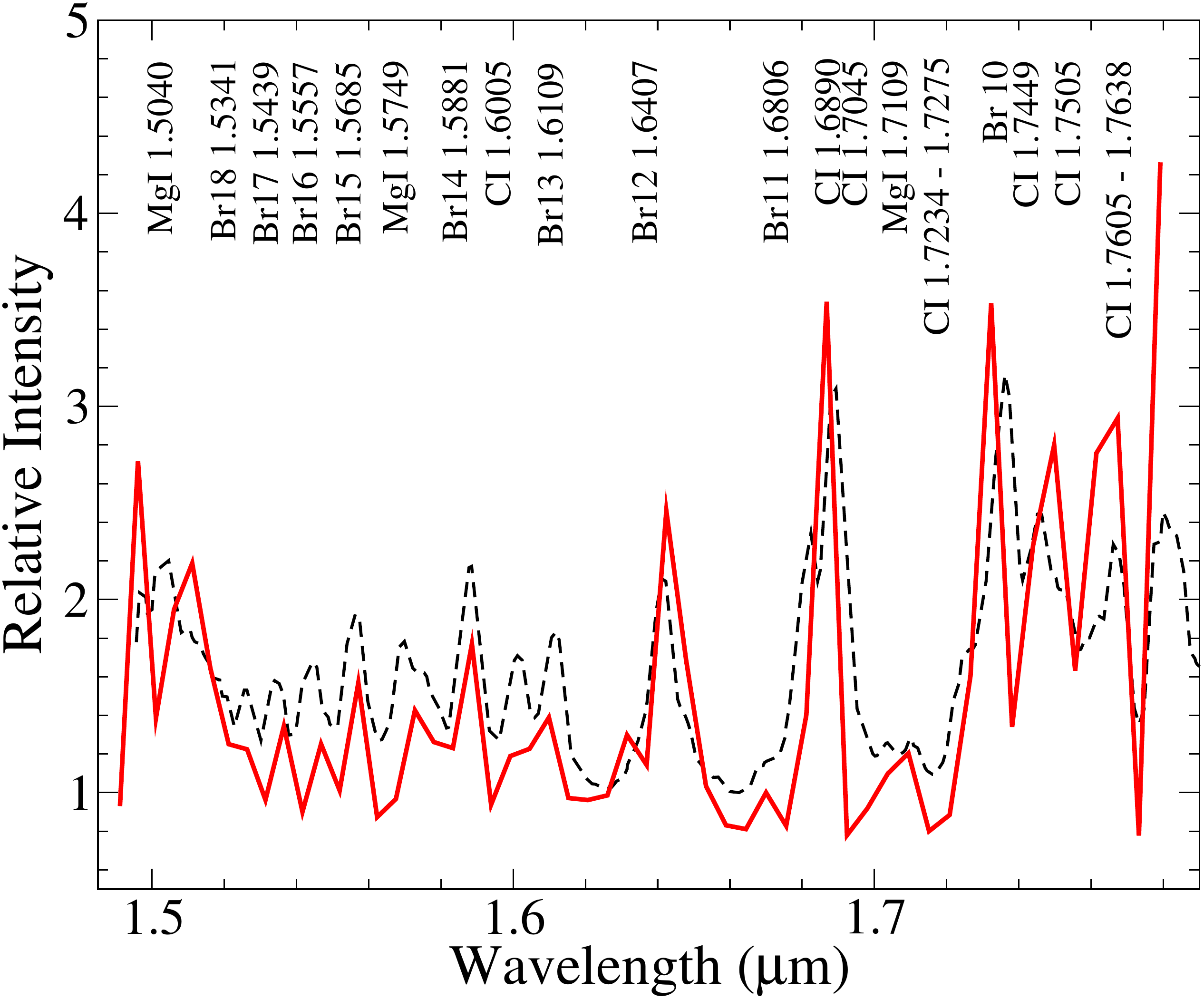}{0.4\textwidth}{(c)}
       \fig{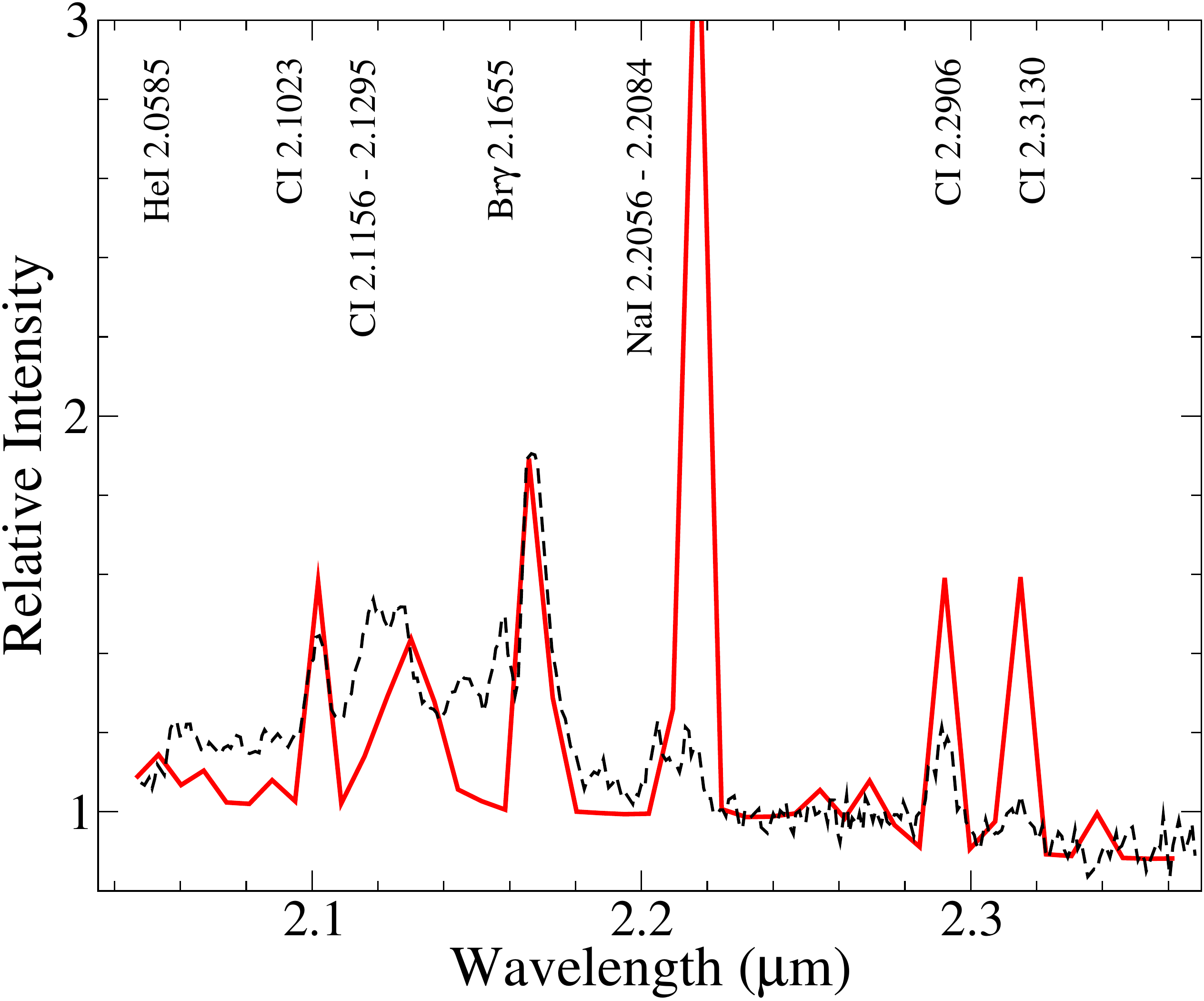}{0.4\textwidth}{(d)}
        }
\caption{Observed (black dashed solid) and model generated spectra (red solid line) of epoch 2 (pre-dust phase). (a) Optical band (the observed optical spectrum is zoomed from $\sim$ 0.54 - 0.65 $\mu$m to show the weak features). (b) J-band. (c) H-band. (d) K-band. For more details see section~\ref{results}.
\label{fig:ep02}}
\end{figure*}

\begin{figure}[ht!]
\gridline{\fig{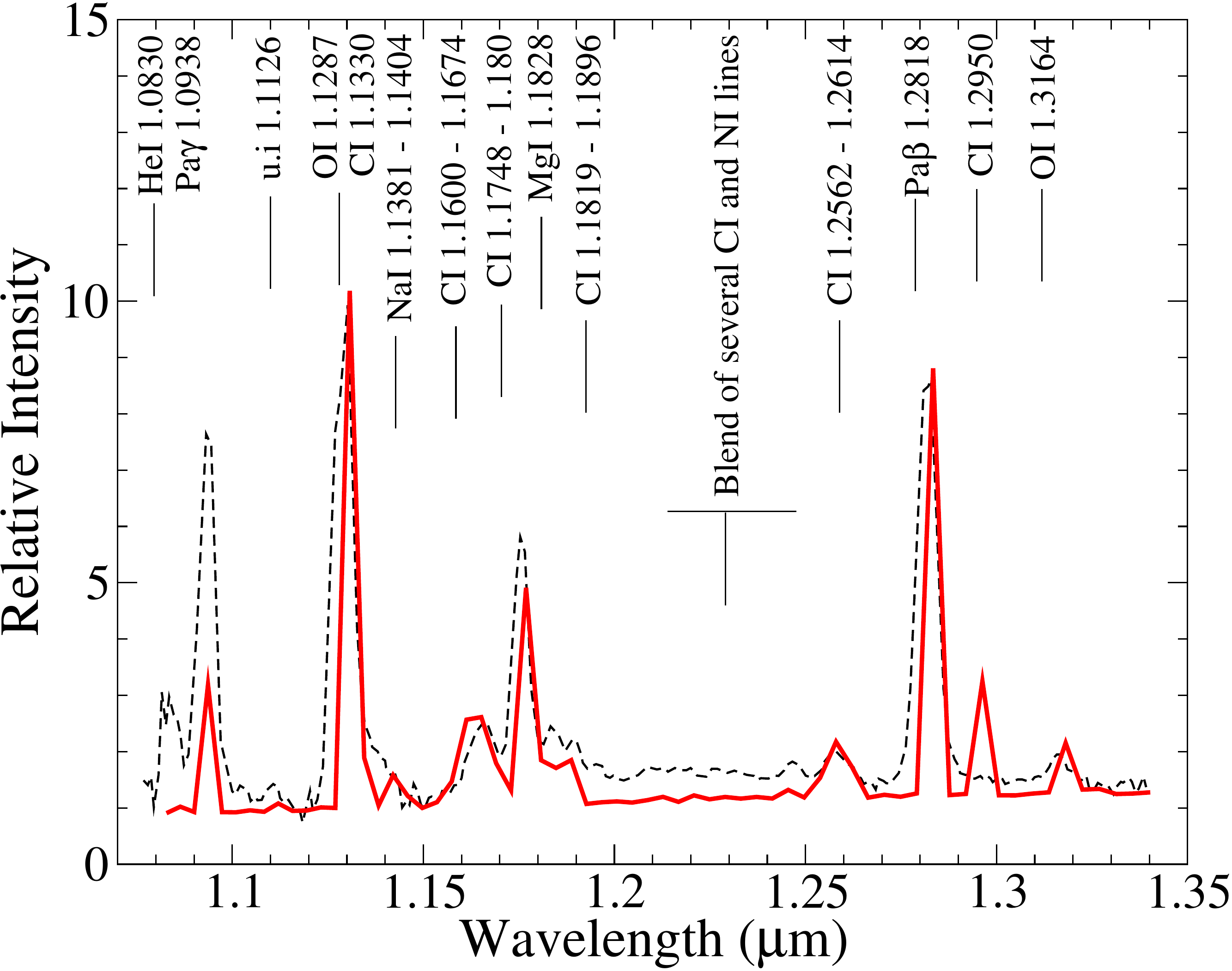}{0.4\textwidth}{(a)}}
\gridline{\fig{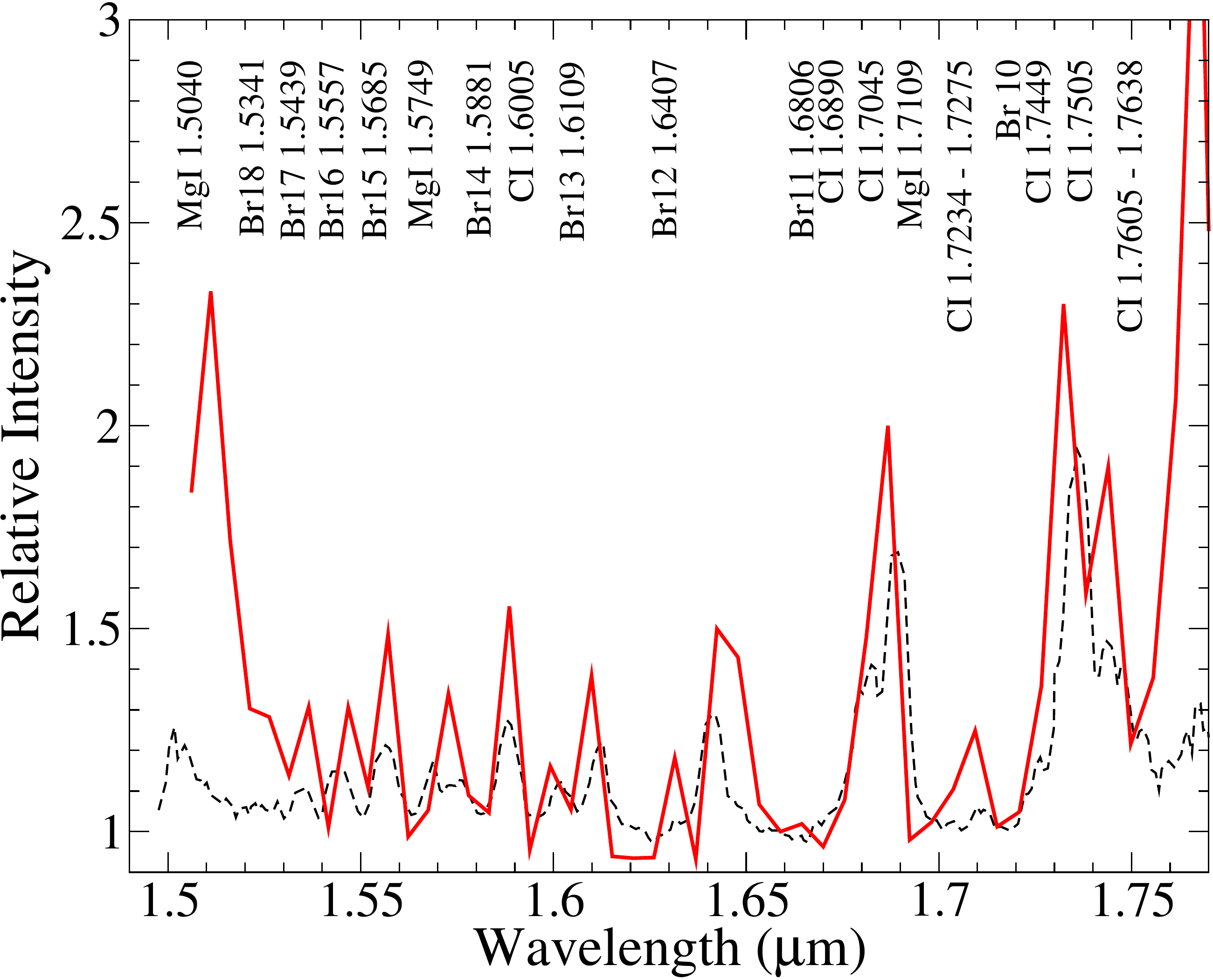}{0.4\textwidth}{(b)}}
\gridline{\fig{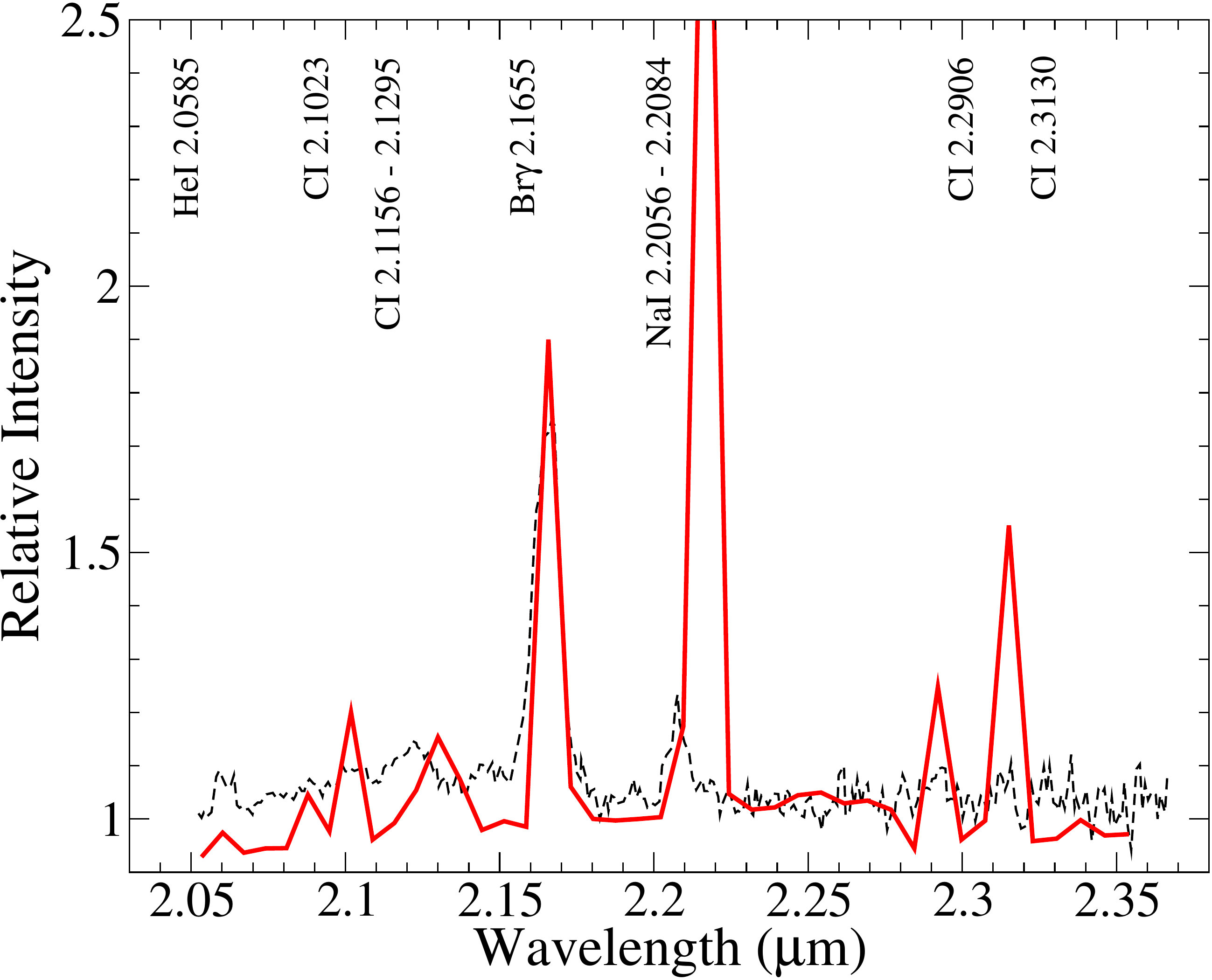}{0.4\textwidth}{(c)}}
\caption{Observed (black dashed solid) and model generated spectra (red solid line) of epoch 3 (post-dust phase). (a) J-band. (b) H-band. (c) K-band. For more details see section~\ref{results}}
\label{fig:ep03}
\end{figure}

\begin{figure}[ht!]
\gridline{\fig{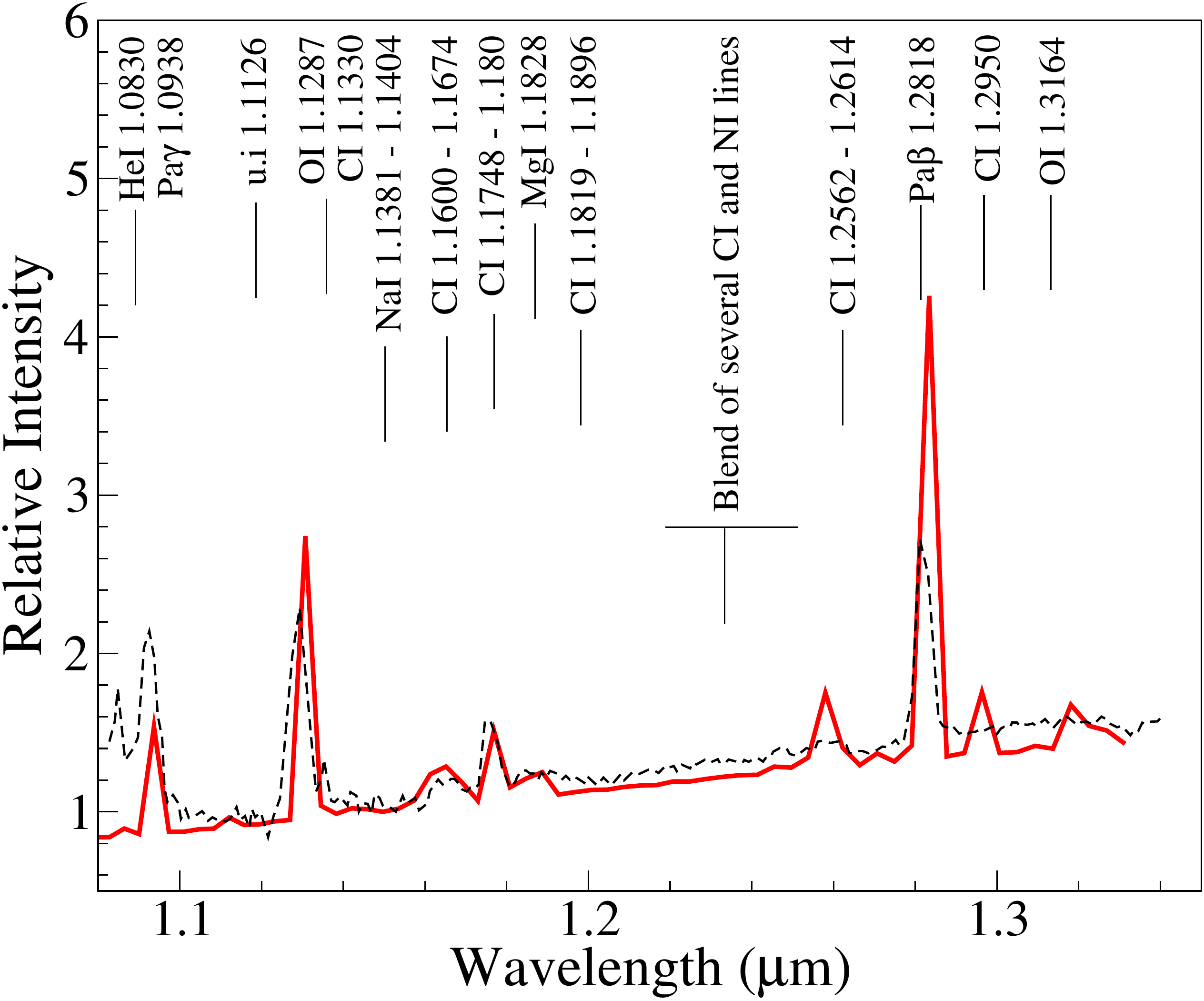}{0.4\textwidth}{(a)}}
\gridline{\fig{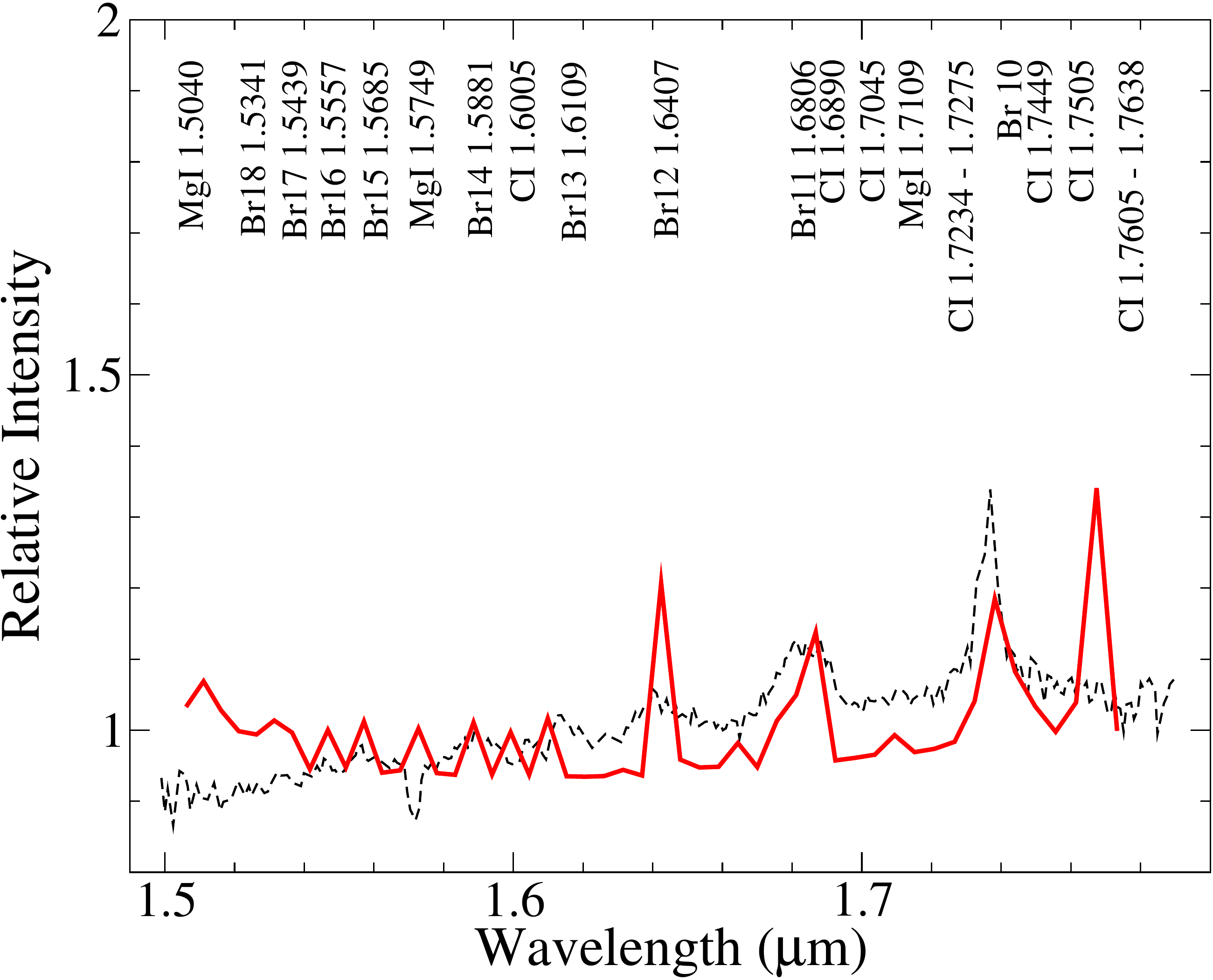}{0.4\textwidth}{(b)}}
\gridline{\fig{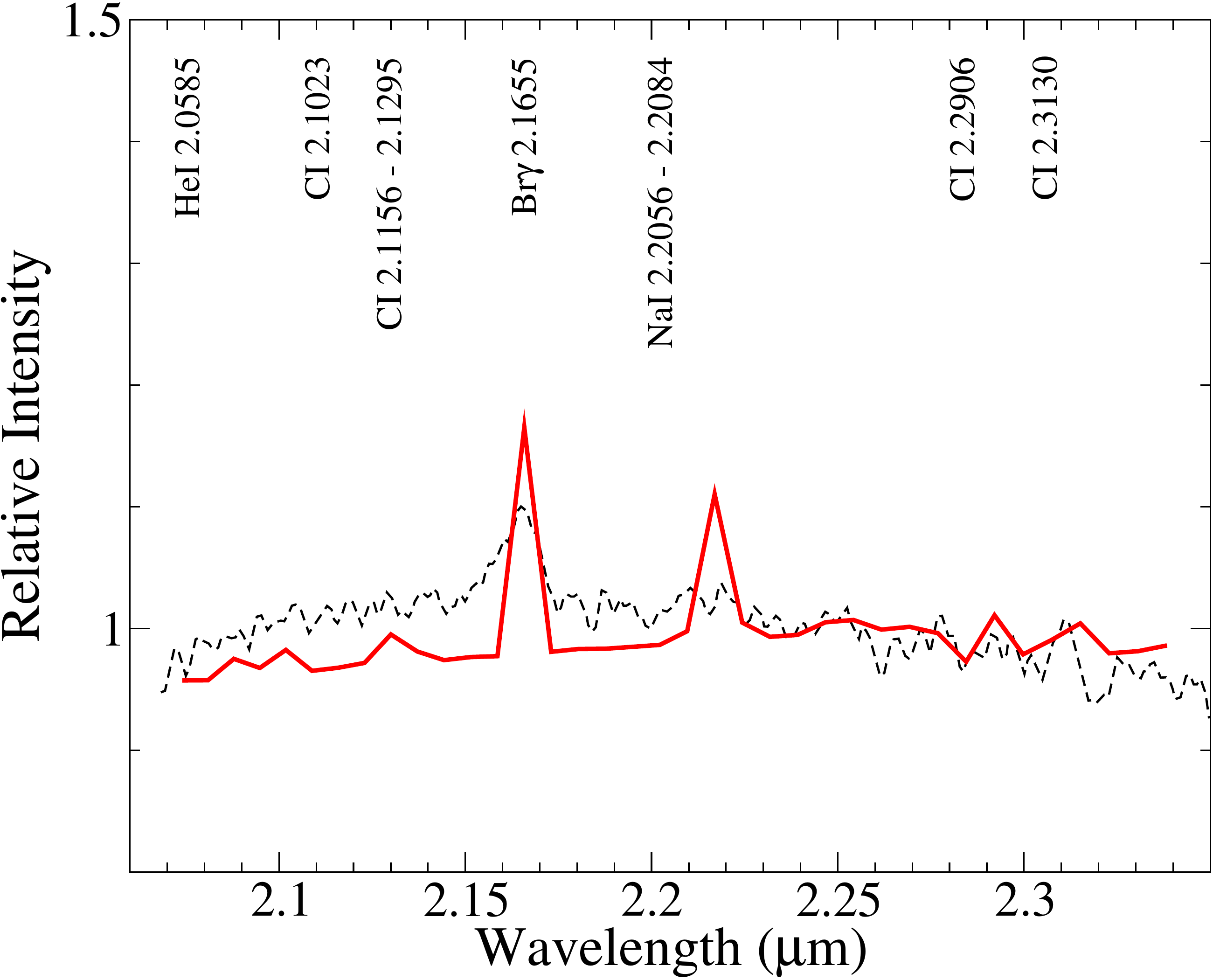}{0.4\textwidth}{(c)}}
\caption{Observed (black dashed solid) and model generated spectra (red solid line) of epoch 4 (post-dust phase). (a) J-band. (b) H-band. (c) K-band. For more details see section~\ref{results}}
\label{fig:ep04}
\end{figure}

\section{Results \& Discussions}\label{results}
The model results for the best-fit \textsc{cloudy} parameters for pre-dust phase and post-dust phase are presented in Table~\ref{tab:fit}. Relative fluxes of observed emission lines, best-fitted \textsc{cloudy} modeled lines, and their corresponding $\chi^2$ are presented in Table~\ref{tab:flux}. While calculating the $\chi^2$ values, we have considered only those lines that are present both in the model-generated and observed spectra. The observed emission lines are dereddened using E(B-V) = 0.4 (Naito et al., 2012; Das et al., 2008) and matched with the spectra generated by \textsc{cloudy}. The \textsc{cloudy} computed best-fit spectra (red solid line), overlaid with observed optical and NIR spectra (black dashed lines) for epochs 1, 2, 3, and 4, are presented in  Fig. \ref{fig:ep01}, \ref{fig:ep02}, \ref{fig:ep03}, and \ref{fig:ep04}, respectively, for comparison. The prominent emission lines are marked in the figures. The observed line fluxes have been determined using \textsc{iraf}\footnote{IRAF is a collection of software written at the National Optical Astronomy Observatory geared towards the reduction of astronomical images in pixel array form. \url{http://ast.noao.edu/data/software}} tasks; the blended features have been fitted with multiple Guassians using \textit{deblend} task in \textsc{iraf} to check the strengths of each line features. The modeling procedure of the pre and post-dust phase spectra has been described in section~\ref{cloudy} in detail.  To minimize errors associated with flux calibration at different wavelengths, the modeled and observed flux ratios have been calculated relative to H$\beta$ in the optical region, Pa $\beta$ in the J band, Br 12 in the H band, and Br $\gamma$ in the K band. The $\chi^{2}_{\text{red}}$ values are 1.90, 1.95, 1.83, and 1.67 on epoch 1, 2, 3, and 4, respectively. The low values of $\chi^{2}_{\text{red}}$ indicate that the fits are satisfactory. From the best-fit model, we estimate the values of different parameters. During modeling, we note that the spectral features are very sensitive to few input parameters, and the features change significantly with even small changes in those parameters. We discuss these below in detail.

\subsection{Temperature and luminosity}
From the best-fit results, we find the values of temperature of the central ionizing source to be $1.32 \times 10^4$, $1.50 \times 10^4$, $1.58 \times 10^4$, and $1.62 \times 10^4$ K, for epochs 1, 2, 3, and 4, respectively. The luminosity is determined to be $2.95 \times 10^{36}$, $3.16 \times 10^{36}$, $3.23 \times 10^{36}$, and $3.31 \times 10^{36}$ ergs$^{-1}$, for epochs 1, 2, 3, 4, respectively. The derived blackbody temperature and luminosity of the central ionizing source increase from epoch 1 to epoch 4. The Lyman $\beta$ pumping increases the strength of O I line at 1.1287 $\mu$m from epoch 1 to epoch 2, which is best fitted by a relatively higher value of temperature. The temperature of the central ionizing source increases possibly because the pseudo photosphere collapses onto the surface of WD (Bath \& Shaviv 1976). At higher temperatures, more radiation is emitted at higher energies than the Lyman limit which could be absorbed by the neutral hydrogen (Bath \& Harkness 1989; Williams et al. 2013) and help in the formation of dust nucleation sites.

\subsection{Ejecta density}
The best-fit spectra are obtained for the density of $1.41 \times 10^{11}$ and $1.77 \times 10^{11}$~cm$^{-3}$, of diffuse part for epochs 1, and 2, respectively. Most of the emission lines are generated successfully from the diffuse part of the ejecta. However, the lines of MgI (1.1828$\mu$m, 1.5040$\mu$m, 1.5749$\mu$m, and 1.7109$\mu$m), NaI (2.2056$\mu$m and 2.2084$\mu$m), are generated from the clump part of the ejecta, where the hydrogen density is high. The values of hydrogen density for the clump part of the ejecta from the best-fit model is found to be $7.94 \times 10^{13}$ and $3.54 \times 10^{13}$~cm$^{-3}$, for epochs 1, and 2, respectively. The Fe II lines are originated from the clump part of the ejecta with a higher hydrogen density. The blended feature of  C I lines in the region between 1.1600-1.1896$\mu$m, and the blended O I 1.1287 $\mu$m and C I 1.1330 $\mu$m line, whose strength increases from epoch 1 to epoch 2, are best fitted by a relatively higher value of hydrogen density of the diffuse part of the ejecta. If the hydrogen density of the diffuse part of the ejecta is decreased, the relative strengths of these lines become weak. We also tried to match these lines by changing the abundances, but we noticed that these observed features of C I are best fitted only at higher hydrogen density and temperature in epoch 2. 

During post-dust phase, from the best-fit \textsc{cloudy} model, we find the values of hydrogen density of clump part to be $3.46 \times 10^{13}$ and $1.00 \times 10^{13}$~cm$^{-3}$, for epochs 3, and 4, respectively. The density of the diffuse part is determined to be $2.23 \times 10^{11}$ and $6.07 \times 10^{11}$~cm$^{-3}$, for epochs 3, and 4, respectively. The existence of such a high-density region within novae ejecta has also been indicated by other studies (e.g., Rawlings 1988;  Rawlings et al. 1989; Pontefract \& Rawlings 2004; Derdzinski et al. 2017) in the dust forming novae. Moreover, the observed spectra of V1280 Sco show the presence of Na I and Mg I, whose ionization potentials are low (5.13eV and 7.64eV, respectively). This implies the presence of a cooler zone within the ejecta which could be associated with the presence of cool clumpy dense sites needed for the formation of dust and molecules in the novae ejecta (Das et al., 2008; Banerjee \& Ashok 2012).

\subsection{Elemental abundances}
The elemental abundances are determined to be He =1.2, C = 13.5, N = 250, O = 27, Na = 1.4, Mg = 1.2, and Fe = 1.0, for epoch 1, He =1.2, C = 20, N = 250, O = 35, Na = 1.3, Mg = 1.3, and Fe = 1.0, for epoch 2, during pre-dust phase, while the abundances has decreased in post-dust phase to He =1.0, C = 15, N = 250, O = 33, Na = 1.3, Mg = 0.5, and Fe = 1.0, for epoch 3, and He =1.0, C = 8, N = 250, O = 28, Na = 1.3, Mg = 0.5, and Fe = 1.0, for epoch 4, by number relative to hydrogen and relative to solar. The elemental abundances determined from our best-fit model with associated uncertainties are presented in Table~\ref{tab:fit}. We do not estimate uncertainties for He, N, and Fe, since only a few lines of He and N were observed in spectra at lower strengths. The best-fit model predicts enhancement in the abundances of He, C, O, N, Na, and Mg, relative to solar values. N was found to be heavily enhanced relative to solar. Also, numerous prominent emission lines of neutral C are present in NIR spectra of V1280 Sco. We used the fittings of these emission lines to determine the C abundance. We notice that the strong H-band C I emission lines in the region 1.7449 and 1.7814$\mu$m are very sensitive to the input parameters. Even a small variation in any of the input parameters causes a significant change in the relative intensities of these emission lines. These lines could be generated at very low C abundance, but, low values of C abundance fails to generate the prominent C I blended features in the region between 1.1600 to 1.1896 $\mu$m in the J band. We observe that a high value of C  abundance (C/H $>$ 15) is required to generate all the C I lines present in the observed spectra.

The O abundance is determined from the spectral fit of the prominent J band O I lines: the 1.1287$\mu$m O I line which increases due to an increase in the Ly $\beta$ flux with time as the temperature of the central ionizing source increases, and the 1.3164 $\mu$m O I line which is excited by a combination of continuum fluorescence and collisional excitation (Banerjee \& Ashok 2012). The mechanism of fluorescence pumping is included as excitation processes for all lines in \textsc{cloudy}. These J-band OI features are reproduced well only for the high value of O abundance. The higher value of oxygen abundance overestimates the optical [O I] emission lines at 5577,  6300, and 6364  \AA~ which are seen at a very low strength (see Figure~\ref{fig:ep02}) on Epoch-02. However, the models with lower O abundance can not fit the J-band O I features. We also tried to generate the optical [O I] lines by varying the other parameters, e.g. T$_{BB}$, L, and $n(H)$ simultaneously with the O abundance, but those combinations could not fit all oxygen features. Also, they disturbed the fitting of the entire NIR $JHK$ spectra, especially the CI features. Hence, we emphasized on fitting the J-band OI features to estimate the oxygen abundance. Our model predicts that O is more abundant than C despite the presence of carbon-rich dust in the ejecta. This is in line with the theoretical results which predict that O is more abundant in the gas phase (Starrfield, Iliadis \& Hix 2016), and it is possible to form C-rich grains in an environment with O $>$ C. The theory of TNR also suggests that O $>$ C irrespective of the type of the white dwarf. 
 
In the region between 1.2 and 1.275 $\mu$m, a very complex blend feature of several C I and N I lines is present at very low strength. Our model generated this feature, but the strength is very low and it's difficult to identify the lines and measure their fluxes in this region. We tried to reproduce this region by trying all the possible combinations of input parameters. We find that this region is generated better only for the higher N abundance. The region is produced best for the N abundance of about $\sim$ 250. The large value of N abundance predicted by our model is the result of proton capturing during TNR (Starrfield, Iliadis \& Hix 2016). The theory of TNR also predicts the abnormal elemental abundance enhancement of heavy elements (particularly CNO) relative to their solar values, in the novae ejecta (Gehrz et al. 1999). Such a high abundance in novae ejecta has also been reported in the case of other novae, e.g., C  = 24, N = 320, O  = 24 (Nova Aql, Snijders 1984); N=316, O=26, Fe=1.0 (Nova Cygni 2006, Munari et al., 2008); He = 1.6, N = 144, O = 58 (V1065 Cen, Helton et al., 2010). Hauschildt et al (1994) reported an enhancement of CNO abundance by a factor of 100 for the dusty nova V705 Cas. A detailed discussion on the abundances in novae ejecta is present in Gehrz et al. (1998), Jos\'e \& Shore (2008). High abundances of the ejecta influence the thermal equilibrium of the gas which might create a favorable environment for the nucleation sites because enhanced metallicity cools the ejecta efficiently and the thermal equilibrium is attained at lower kinetic temperature (Ferland \& Shields 1978). Shore et al. (2018) also discussed that a very large value of N/C, relative to solar could also play a significant role in the formation of dust in CO novae.

\subsection{The Fe II feature}
The optical spectrum on epoch 2 is dominated by numerous emission lines from Fe II along with Balmer lines, and low-ionization lines. This optically thick stage of nova is known as \enquote{iron-curtain} phase in which the nova ejecta is relatively cool. During this phase, the UV spectrum shows a complex overlapping band of absorption lines from low ionization metals, and most of the opacity is contributed by the line blanketing from Fe lines (Shore et al. 1994; Shore et al. 2016). We used this epoch to fix the value of Fe abundance. The Fe II multiplets between 4500 \AA~and 5500 \AA~ are fitted well for higher hydrogen density ($\ge 10^{13.0}$ cm$^{-3}$) of the clump part and a solar value of Fe. Though our model could generate most of the Fe II multiplets well, the two Fe II emissions at 5018 \AA~and 5169 \AA~could not be fitted with the observed line feature.

In the observed J band spectra, a weak feature at 1.1126$\mu$m is present which is reported by Das et al. (2008) as an unidentified line. From our model analysis, we identify this line as a Fe II line which is a collisionally excited line (Rudy et al. (2000). This line belongs to one of the so-called \enquote{1 $\mu$m Fe II lines}, which are present at 0.9997, 1.0501, 1.0863, and 1.1126 $\mu$m in other astronomical objects also (Rudy et al. 1991; Rudy et al., 2000). These Fe II emissions have also been observed in the spectra of a few novae in NIR domain (Das et al., 2008; Das et al., 2009). The main excitation mechanism for these Fe II emissions is primarily by Ly $\alpha$ fluorescence (Banerjee et al. 2009). The other excitation mechanism is considered to be collisional excitation and Lyman continuum fluorescence (Banerjee et al. 2009 and references therein).

\subsection{Dust characteristics}\label{result:dust}
To include dust in the model, first, we added the grains at the inner shell in the two-component model that we used for the pre-dust phase. However, this attempt fails because the inner part of the ejecta is (R$_{in} \sim 10^{14.02}$cm) is very close to the central ionizing source and faces the harsh radiation emitted by it. Consequently, the gas temperature is very high in comparison to the condensation temperature $\sim$1300K, (Evans 2001) of the dust grains. For a very high temperature, the molecules would not form due to the collisional dissociation. The rates of collisional dissociation reduce once the gas temperature reaches a lower value ($<$ 4000 K) (e.g., Roberge \& Dalgarno 1982; Rawlings \& Williams 1989; Pontefract \& Rawlings 2004, etc), and the possibility of dust formation increases. Pontefract \& Rawlings (2004) used their chemical model including 1110 chemical reactions of some 76 species, which are composed of the elements C, N, O, H, He, and Na, and showed that most species demonstrate a very strong dependence on the ejecta density, and even for a small molecule formation, the ejecta density must be very high. In a previous study Rawlings (1988) showed that a high density ($\sim$ 10 - 100 times the mean value) is required for the formation of dust nucleation sites and to protect the dust molecules from external harsh radiation environment. Rawlings \& Williams (1989) showed that the dust formation chemistry in harsh nova ejecta highly depends on the amount of hydrogen in the form of H$_2$, densities higher than some critical value do not allow the radiative stabilization of vibrationally excited H$_2$. Molecules could condense to form dust grains in an environment that has predominantly neutral gas and is sufficiently cool. This condition in the ejecta may be achieved during \enquote{iron-curtain} phase when the UV transitions are thick which makes the outer region of the nova ejecta cold enough to allow the formation of molecules. Internal shocks in the nova ejecta could also lead to the condensation of dust grains. Strong radiative shocks via compression could create a cool and dense environment ($\geq 10^{14}$cm$^{-3}$) which results in saturation of CO and rapid dust nucleation (Derdzinski et al 2017). In their shock model, Derdzinski et al. (2017) suggested that in the post-shock regions dust grains can form and survive from the intense UV and X-rays radiations from the central WD. They proposed that these high-energy particles accelerated at the shock could destroy CO molecules, allowing the formation of carbon dust in post-shock cool and dense regions. More detailed discussion about shocks in novae ejecta can be found in Steinberg \& Metzger (2020); Chomiuk et al. (2019); Metzger et al. (2014), and references therein. Considering all the aspects for the dust formation in nova, it is clear that the dust grains could not be added at R$_{in}$ in our model, instead grains can be added to the outer layers of the clump ejecta where the temperature is low and density is still sufficiently high.

To determine the necessary temperature and hydrogen density, we check the calculations of the clump component at different gas temperatures in the range of 1000 K to 4000 K.  We find that when the temperature of ejecta is around 2000K, the hydrogen density is calculated using Eq \ref{eq:one} at a radius of ($\sim 4.07\times 10^{15}$ cm), is still high enough ($3.16 \times 10^{8}$ cm$^{-3}$) which allows the dust formation. A similar value of gas temperature (2000 K) and gas density ($5 \times 10^9$cm$^{-3}$) is used in the theoretical work by Rawlings \& Williams (1989). Shore et al. (1994) also showed that dust must form in the outer part of the ejecta at a radius of the order $\sim 10^{15}$cm, where the temperature is low and the density is sufficiently high. Also, the existence of neutral carbon is required for the formation of molecules (Rudy et al. 2003; Bode \& Evans 2008). Carbon remains neutral until the density decreases to $\sim 10^{9} - 10^{10}$ cm$^{-3}$, and these values are achieved at gas temperature, T $\ge$ 2000 K (Derdzinski et al. 2017).  Considering the results reported by various authors as discussed earlier in this paper, the idea to add dust in the outer region of the ejecta where the gas temperature is $\sim$ 2000 K and radius is $\sim 4.07\times 10^{15}$ cm seems a reasonable choice. The final spectrum is obtained by multiplying the spectrum from each component with their corresponding covering factors and then adding them.

Dusty novae are known to produce copious amounts of amorphous carbon and silicate (e.g., V1065 Cent, Helton et al. (2010); V2676 Oph, Kawakita, et al. (2017);) and some also produce amorphous carbon, SiC, silicates, and hydrocarbons altogether (e.g., V842 Cen (Gehrz 1990); QU Vul, Gehrz et al. 1992), etc. Dust formation in novae could proceed under non-equilibrium conditions when only a small fraction of available carbon ends up in CO formation and the carbon and oxide grains can condense simultaneously (Bose \& Starrfield 2019). Jos\'{e} et al. (2004) showed that the presence of intermediate-mass elements, such as Al, Ca, Mg, or Si, could lead to these non-equilibrium conditions by altering the condensation process which further allows the formation of carbon dust in a slightly oxygen-rich environment. Moreover, Shore \& Gehrz (2004) showed that the strong radiations from the central ionizing WD cause accretion of grains clusters through induced dipole interactions which promote the formation of dust nucleation sites in novae kinetically rather than at equilibrium. The timing of the C-rich and O-rich grains formation could be different. In several studies of novae the C-rich dust grains were identified first, followed by O-rich dust grains (see e.g., Gehrz et al. 1992; Sakon et al. 2016).  Observations in IR (Evans et al. 2017) and UV (Shore et al. 1994) showed that the size of dust grains in novae ejecta could be $>$ 0.2$\mu$m. In the previous studies of V1280 Sco, from spectral fitting Chesneau et al. (2008) estimated sizes of the dust grains in the range of 0.03-3.0$\mu$m, and $q=2.1$ for the initial stage of dust formation and a slightly larger value of $q \sim$ 3 for the later stages of dust. From this, it appears that the dust grains sizes started decreasing after their formation. Helton et al. (2010) found $q = 3.0$ with the grains size distribution in the range of 0.005 - 5.000 $\mu$m for the dusty nova V1065 Cen. Sakon et al. (2016) incorporated both amorphous carbon of small size 0.01 $\mu$m and astrophysical silicates of large size 0.3-0.5 $\mu$m in their study for V1280 Sco. Following these findings, for simplicity we consider only dust grains of amorphous carbon and astrophysical silicates in our model. The default available grains (e.g. ISM grains, Orion grains, graphite, PAH, silicate, etc.) in \textsc{cloudy} don't resemble novae grains. However, \textsc{cloudy} allows creating dust opacities using built-in refractive indices. \textsc{cloudy} assumes grains as homogeneous spheres and uses a spherical Mie code (Hansen and Travis 1974; van Hoof et al. 2004) to calculate the sets of absorption and scattering opacities (for grain size $<$ 3 $\mu$m) and user-defined size distributions using \textit{compile grains} command. For grains size distributions we construct opacity file for amorphous carbon, and astrophysical silicate for $q=2.1$, which is also in the range of the value ($q = 2.3 \pm 0.5$) as reported by Evans et al. (2005) for the dusty nova V705 Cas. Refractive index data for astrophysical silicate, and amorphous carbon, are taken from Martin \& Rouleau (1991), and Rouleau \& Martin (1991), respectively. The details of the grain model implemented in \textsc{cloudy} are described in van hoof et al. (2004) and Ferland et al. (2013) (section 2.5). We also find that the combination of large (0.03 - 3.0$\mu$m) astrophysical silicate grains and small (0.005 - 0.25 $\mu$m) amorphous carbon grains are required to generate the observed IR continuum in post-dust phase spectra. We also tried to generate the spectra for different other grain sizes to generate the observed post-dust phase spectra, but they failed to generate the continuum of the observed JHK spectra properly.

The estimated values of T and L of the central ionizing WD are relatively low, which are consistent with the previous results. For example, Sakon et al. (2016) assumed M$_{\text{WD}}$ = 0.6 M$_{\odot}$ and determined the T and L of the central WD on Day 150 to be $\sim$ 2.5 $\times$ 10$^4$ K and 1.6 $\times$ 10$^{4}$ L$_{\odot}$, respectively, which further increased to 1.0 $\times$ 10$^{5}$ K and 1.8 $\times $10$^4$ L$_{\odot}$, respectively during the nebular phase. Naito et al. (2012) reported the appearance of He I 5876, 6678 lines about 571 days after the outburst. This indicated the photospheric temperature was not high enough to produce the lines of elements with high ionization potential. The results of the present study show that all of the necessary conditions required for dust formation as found in the previous studies, viz. a relatively lower temperature and luminosity, high ejecta density, high elemental abundances, especially of CNO, have been available in the nova ejecta, which helped in the formation of thick dust in V1280 Sco.

Here, we have approached the problem phenomenologically and we tried to give an overall estimate of the physical parameters associated with the dust formation in the case of V1280 Sco using observed optical and NIR spectra. It has been established that dust grains are likely to survive even after the decrease in IR luminosity, multi-wavelength high-resolution spectroscopy of the later phases of dusty novae could be a useful tool to understand the nature of dust in novae ejecta.

\begin{figure*}[ht!]
\gridline{\fig{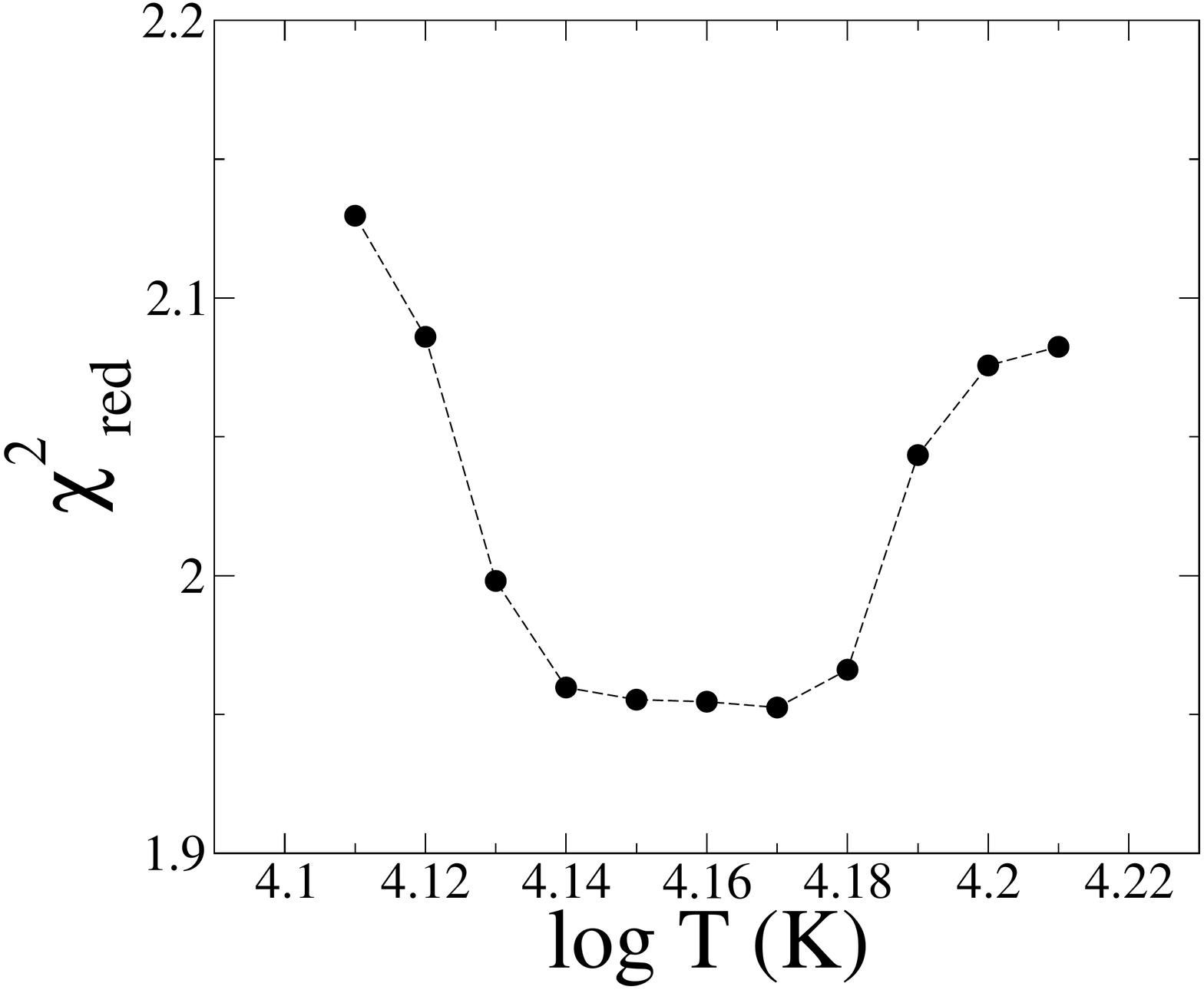}{0.25\textwidth}{(a)}
         \fig{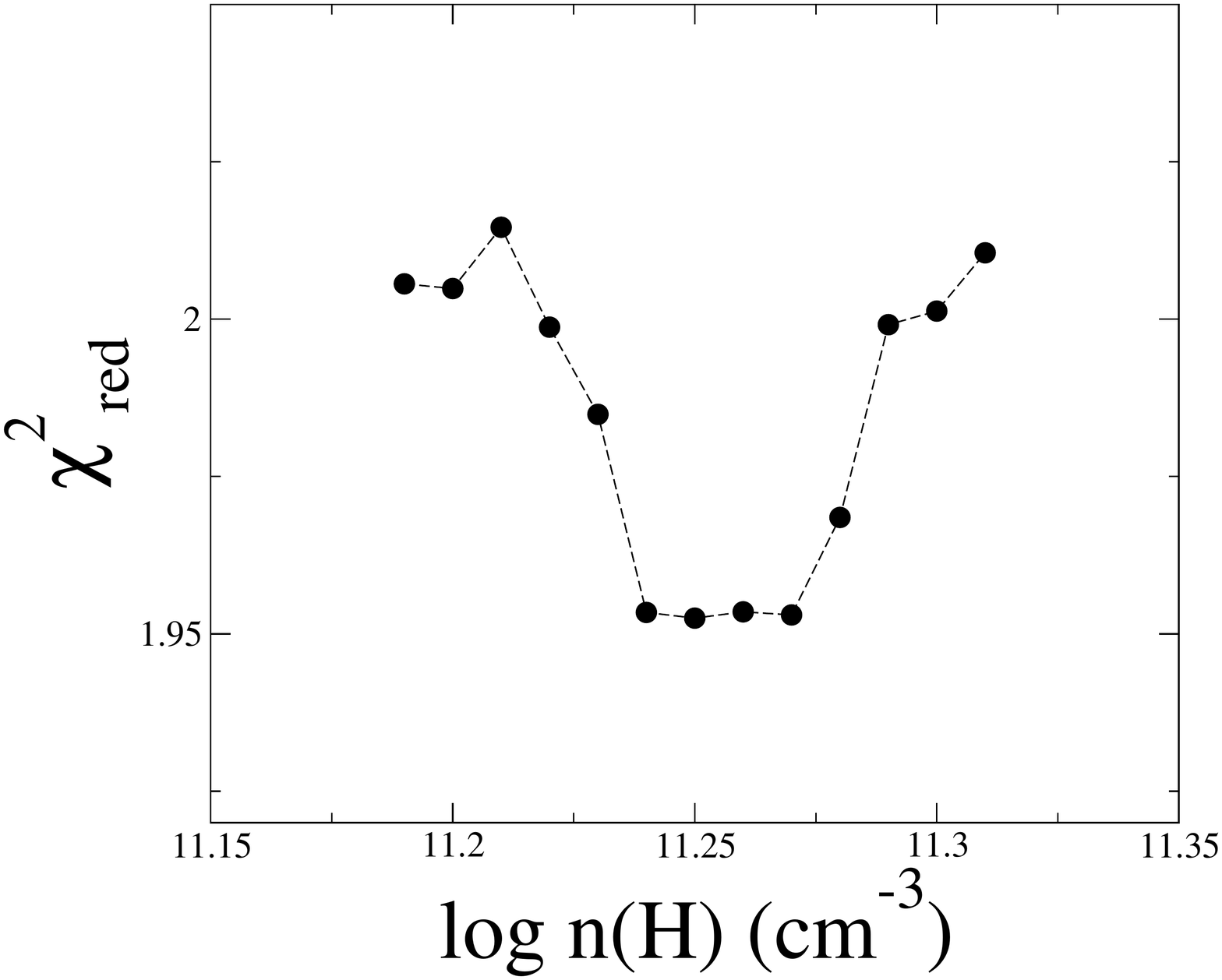}{0.25\textwidth}{(b)}
         \fig{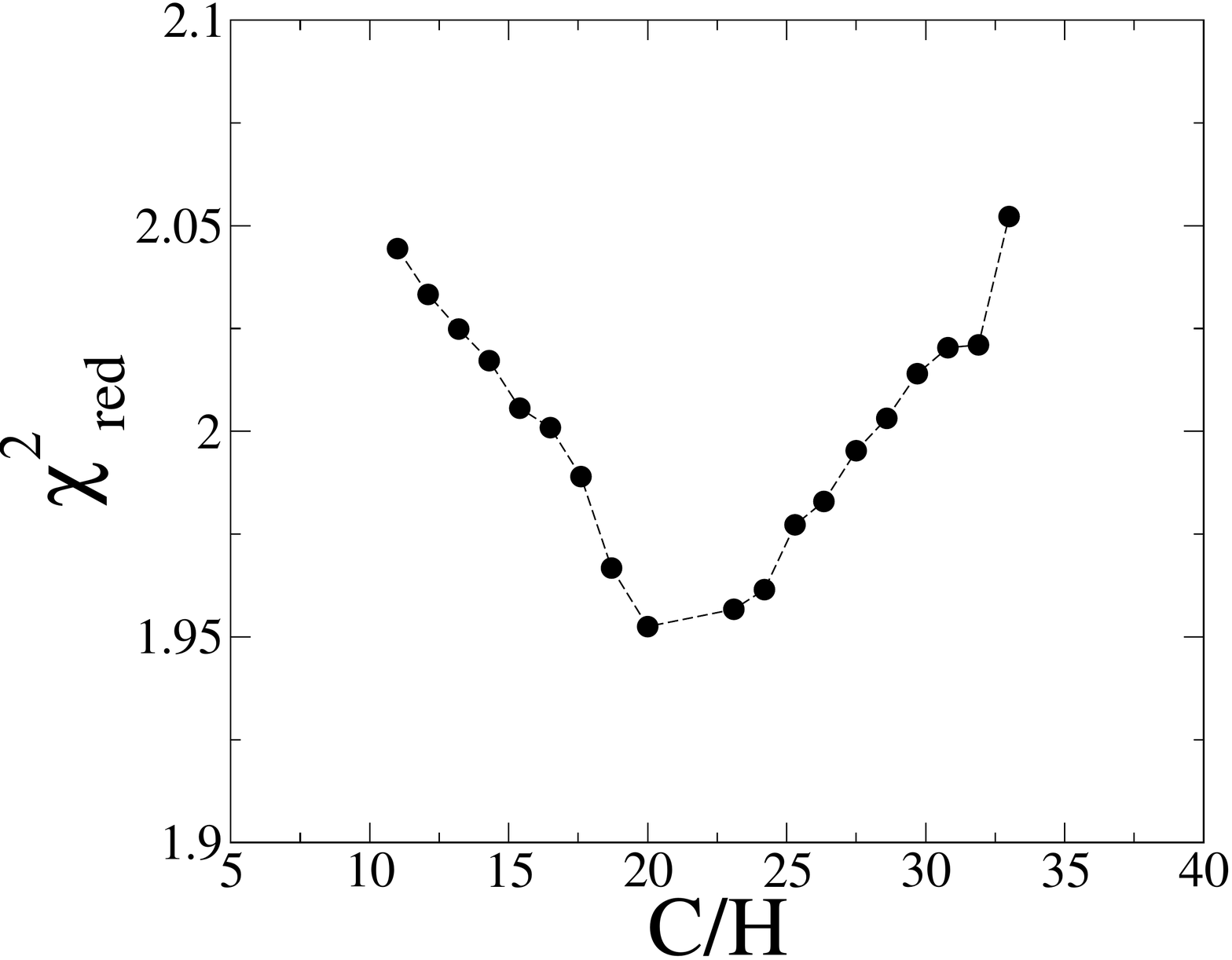}{0.25\textwidth}{(c)}
         \fig{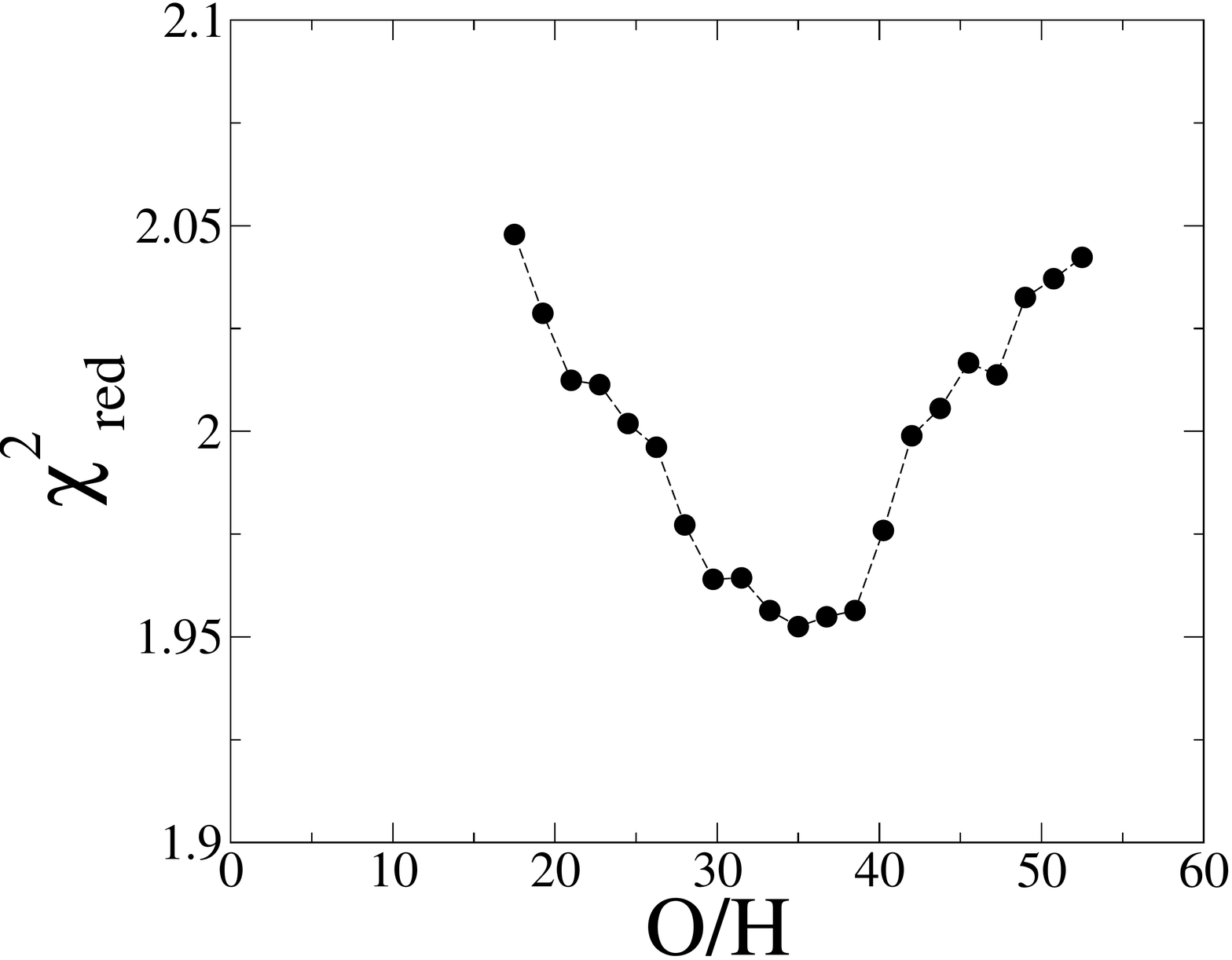}{0.25\textwidth}{(d)}}
\gridline{\fig{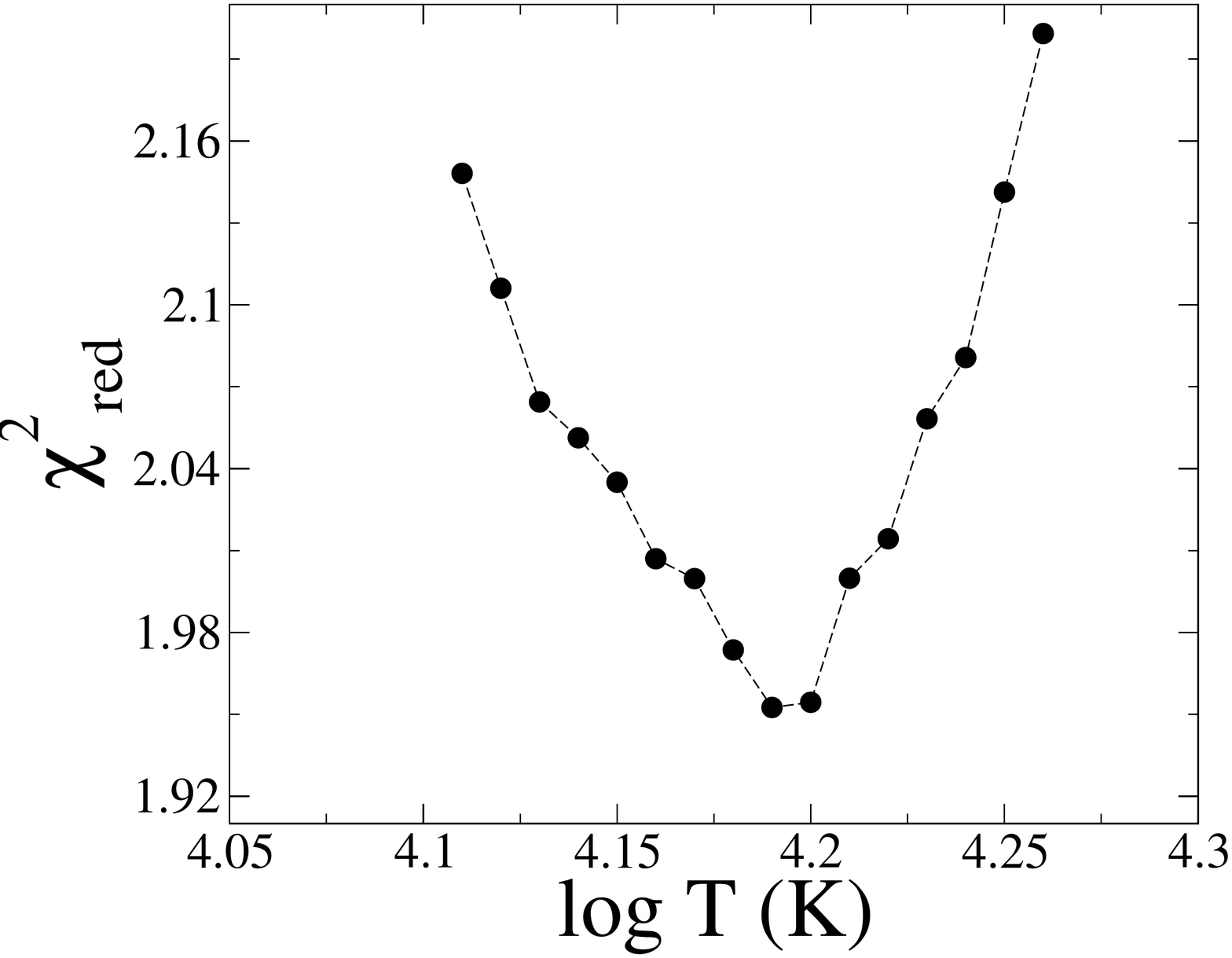}{0.25\textwidth}{(e)}
         \fig{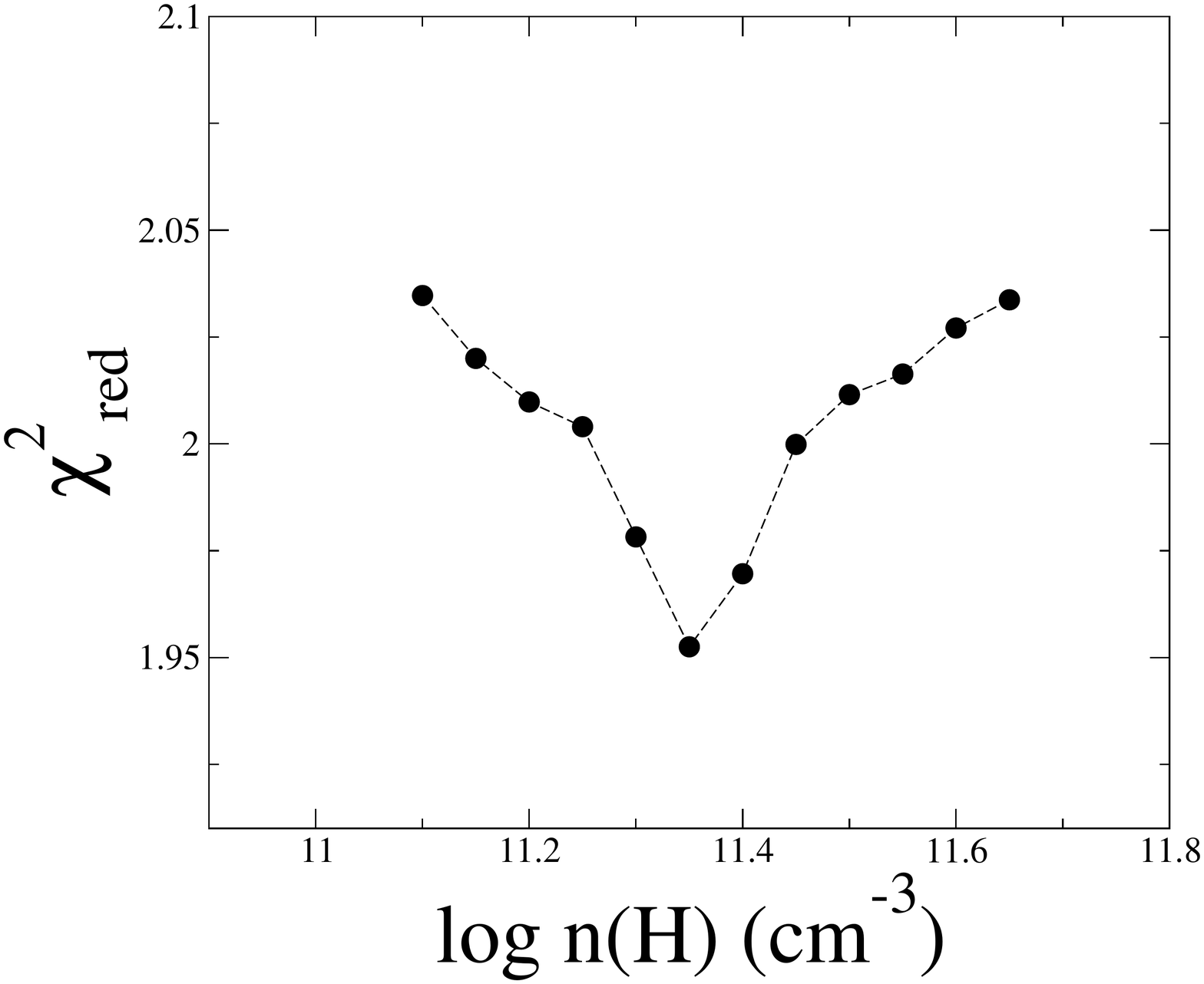}{0.25\textwidth}{(f)}
         \fig{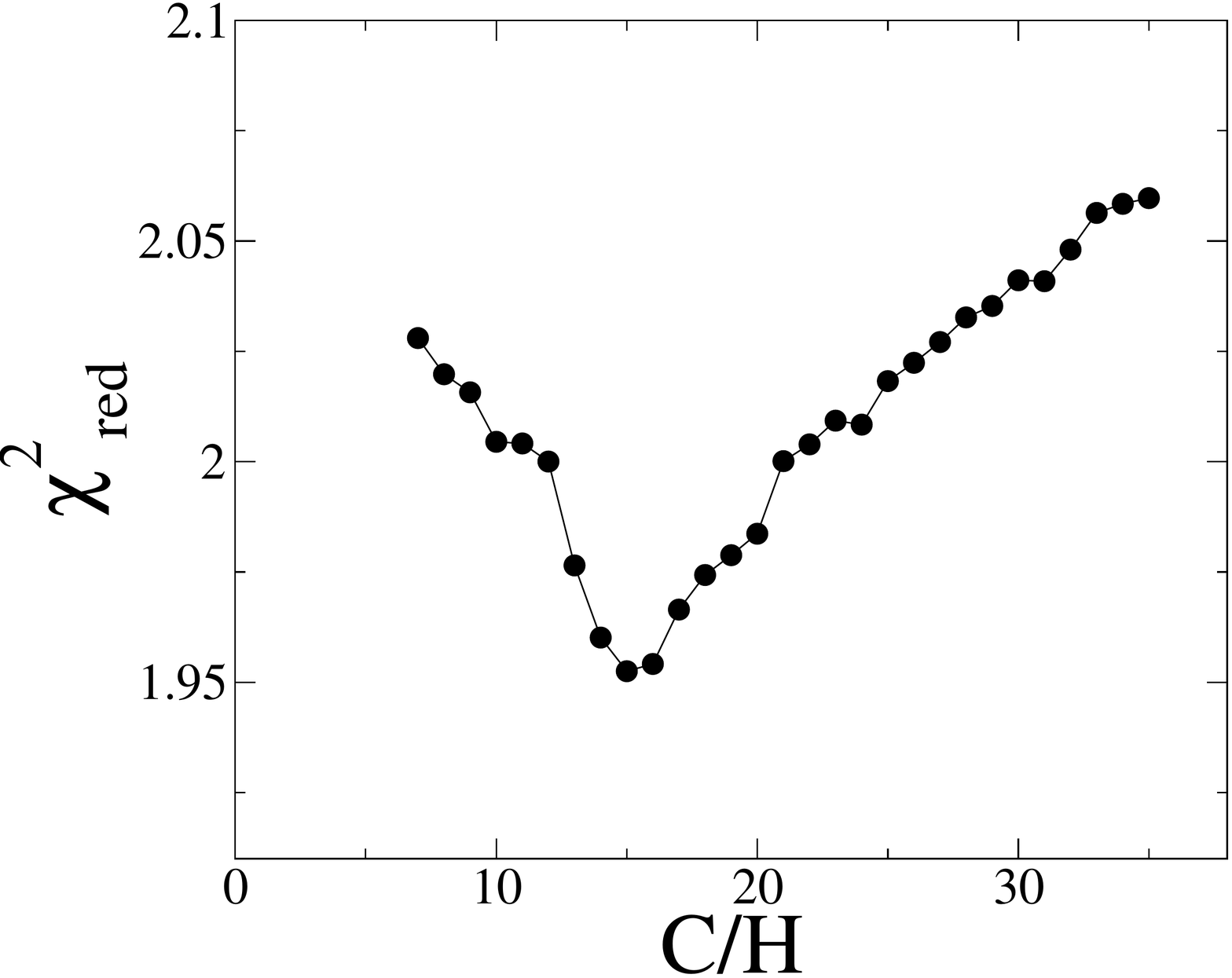}{0.25\textwidth}{(g)}
         \fig{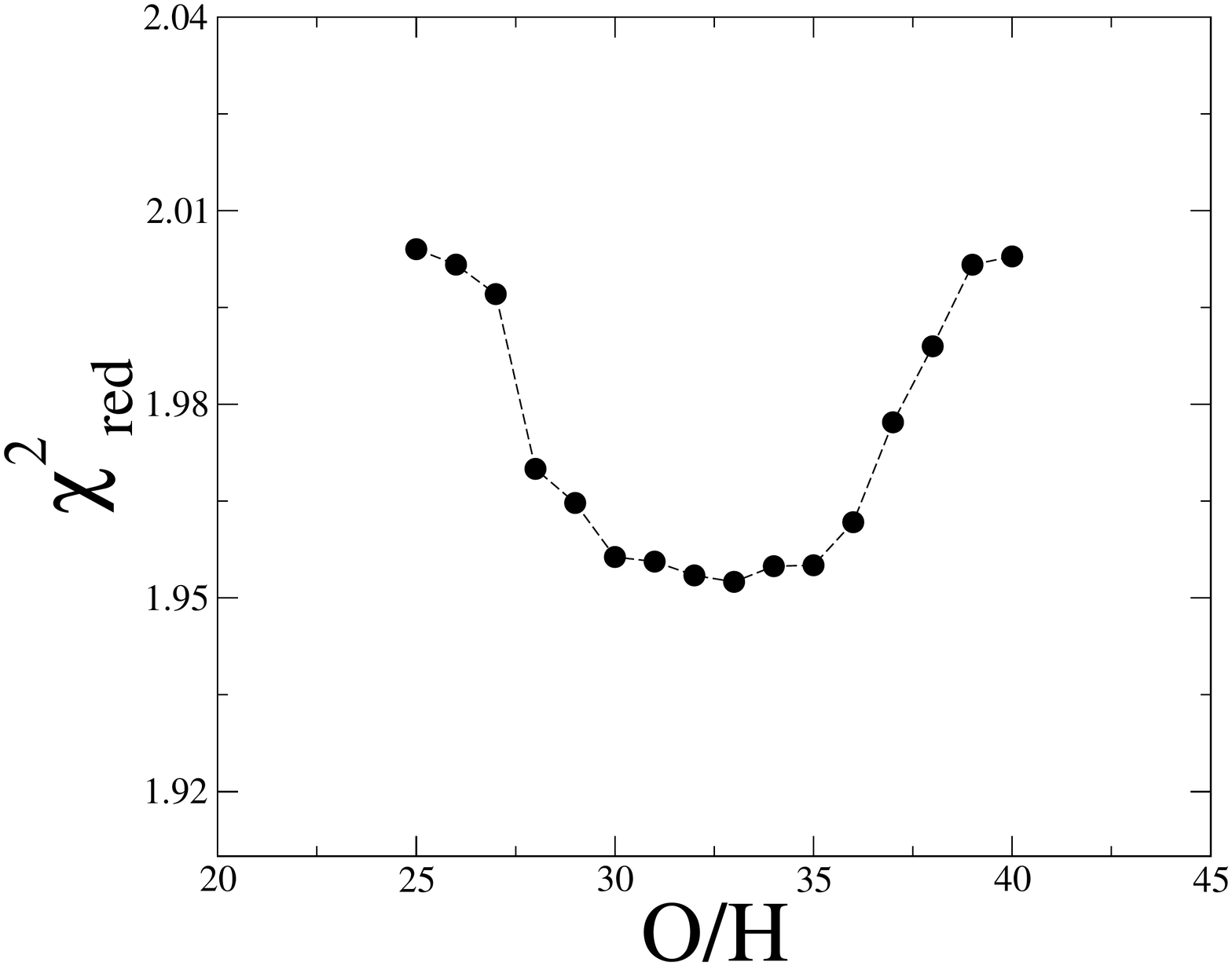}{0.25\textwidth}{(h)}}
\caption{$\chi^2_{red}$ with the model free parameters. Each parameter was varied around their best-fit value to determine the errors in those parameters. For more details see section~\ref{cloudy}. For pre-dust phase (Epoch 2), the variation of $\chi^2_{red}$ with temperatures of central ionizing WD, ejecta density, Carbon, and Oxygen abundance is shown in (a), (b), (c), and (d), respectively. Similarly, for post-dust phase (Epoch 3), the variation of $\chi^2_{red}$ with Temperatures of central ionizing WD, ejecta density, Carbon, and Oxygen abundance is shown in (e), (f), (g), and (h), respectively.
\label{fig:ep07}}
\end{figure*}

\subsection{Model limitations}
The low $\chi_{red}^2$ values indicate that the modeled spectra match well with the observed spectra. Though most of the lines match well, a few lines could not be generated properly. Here, we have assumed a homogeneous spherically symmetric geometry, though, recent studies have revealed the complex nature of the novae ejecta with a morphology (Mason et al. 2018; Chomiuk et al. 2014; Shore et al. 2013; Chesneau et al. 2011; Rupen et al. 2008). The density variations within the ejecta and the non-spherical geometry can be inferred by the saddle-shaped nebular line profiles which are often characterized by fine structures on their top (Shore et al. 2018). Many studies have also used Monte-Carlo-based radiative transfer models to fit the non-spherical geometries of novae (Shore et al. 2003). \textsc{cloudy} being a one-dimensional code, is not well suited to represent this type of complex environment. Our approach to the two-component model may not be perfect, since \textsc{cloudy} handles them separately, but in reality, they are not separate. Presently, \textsc{cloudy} can not be used to add clumpy components within the diffuse component. Moreover, people perform NLTE time-independent snapshot models, in general, to save computational time. However in some cases, at a very early stage, the recombination time is comparable to the age of the ejecta. A dynamical model is a better solution in such conditions. \textsc{cloudy} is capable to do dynamical simulations but it is nearly 1000 times slower than a time-steady calculation. Hence, in this paper, we use NLTE time-independent snapshot models. In the future, our model could be improved by using dynamical simulations.

In the J band, Pa $\beta 1.2818 \mu$m is seen to be weaker than Pa $\gamma 1.0938 \mu$m in the observed spectra, although the model predicts the other way around. We tried different combinations of initial blackbody parameters and density of ejecta, but we couldn't generate the observed line ratio. We also checked one case by considering a constant temperature throughout the ejecta. This generated Pa $\gamma$, and Pa $\beta$ to the observed line ratio, but at the same time, all the C I line features disappear. Our model also predicts a high value of the H$\alpha$ intensity in the optical band. It should be noted that these lines contribute to the large values of $\chi^2$ in the calculations for all epochs. The C I lines in the region between 1.7449 and 1.7814$\mu$m in the H-band is very sensitive to the initial temperature, density, and C abundances. Our model also overestimates the intensity of K band CI lines at 2.2156-2.2167 $\mu$m on all epochs. These lines appear only after adding maximum available energy levels of \textsc{stout} database. The reason behind this is that most optical and infrared emissions are produced by radiative recombination which follows ionization. Highly excited Rydberg levels are populated following recombination from the ion above it, rather than collisional excitation from the ground state, in photoionization equilibrium. Hence increasing the number of energy levels affects the line intensities (for details please check section 2.1 of Ferland et al. 2017). We also observed that the intensity of this line decreases when C abundance is reduced, but the low value of C abundance does not generate other prominent C I features in the J and H band. We also checked the dependency of these C I lines on other parameters. These C I lines are generated for very low and very high hydrogen density. But at those values of hydrogen density, the majority of other lines are not generating properly. Also, our model could not generate He I 1.0830$\mu$m, this line generates at a very high gas temperature (i.e., T $> 10^{4.4}$K), which is not suitable for the generation of other emission lines of relatively lower ionization potential. Hence, these lines (He I 1.0830, C I 1.7449 and 1.7814, 2.2156 - 2.2167$\mu$m) are excluded during the $\chi^{2}$ calculations.

Our model for epoch 4 (post-dust phase) needs further modifications as our approach of modeling the effects of dust grains on the continuum and line intensities is very simple. For the present study, we assume only a mixture of amorphous carbon and astrophysical dust grains, however, in reality, novae are known to produce a mixture of all possible kinds of dust in a single outburst. We are not able to test models with amorphous carbon grains using a size $> 1\mu$m due to insufficient default refractive index data in \textsc{cloudy}. Also, the Mie scattering approximation is not efficient for larger grains with sizes  $>$ 3$\mu$m. A rigorous study is required to investigate the effects of different grain sizes on the observed spectral lines in the near-Infrared domain. The ejecta geometry is also very crucial while studying the detailed structure of the ejecta. The line profiles also depend on the thickness and radial location of the dust within the ejecta. Shore et al. (2018) showed that the line profiles show asymmetries with no change in the maximum velocity if the dust is concentrated in the inner ejecta. Thicker dust decreases the maximum recession velocity with an unchanged blue wing of the line profile. Additionally, for a fixed location of the dust, the line profiles change significantly with the changes in the inclination angle (see Figure 3 from Shore et al. 2018). We will incorporate a more detailed analysis of non-spherical geometry in a future paper.

It may be noted that the approach of determining uncertainties has some drawbacks because this method does not consider the correlations between the free parameters, and thus underestimate the true uncertainties in the model parameters. The number of free physical parameters in our analysis are 13 and 15 for the pre-dust and post-dust phases, respectively. The error estimation in each physical parameter requires a fairly large number of \textsc{cloudy} simulations ($\sim$ $m2^m$, where $m$ is the number of correlated free parameters for each modeled epoch) to fully map the $\chi^2$ space as a function of free parameters, which makes an exact error analysis too expensive computationally. Furthermore, our models can not generate the prominent features of carbon lines using the default \textsc{stout} database which has a lesser number of energy levels. Thus, we use all the available energy levels of the \textsc{stout} database which successfully generate the prominent carbon features between CI $1.1600 \mu$m and CI $1.1896 \mu$m, and the complex blend of NI and CI lines between 1.2 and $1.275 \mu$m, CI $1.2562\mu$m, CI $1.2614\mu$m in J-band, carbon features in the region between 1.7605 and $1.7814 \mu$m in H-band, and the CI lines from 2.2156 to $2.2167\mu$m in K-band (see Fig. ~\ref{fig:ep01}). The use of such a database makes the calculations more complicated and comes at the cost of longer computational time (for more details please see Ferland et al. 2017). Thus, a more precise estimation of errors associated with the parameters is computationally expensive and will take a much longer time. A more correct procedure to estimate the errors associated with the derived values would be to perform a Monte-Carlo analysis from a completely randomized set of all free input parameters varying simultaneously. We plan to perform such kind of detailed error estimation in a future work.

\begin{table*}[h]
\setlength{\tabcolsep}{8pt}
\begin{center}
\footnotesize{
	\caption{Best-fit Cloudy model parameters for V1280 Scorpii}\label{tab:fit}
	\label{tab:predust}
	\begin{tabular}{l l l l l l}
		\hline
		\hline
		Parameters (units) & Epoch-01 (Pre-dust) & Epoch-02 (Pre-dust) & Epoch-03 (post-dust) & Epoch-04 (Post-dust) \\
		\hline
		Temperature ($\times 10^{4}$ K) \dotfill & 1.32${\pm 0.06}$ & 1.50$^{+0.03}_{-0.09}$ & 1.58${\pm 0.01}$ & 1.62$^{+0.03}_{-0.01}$  \\
		Luminosity ($\times 10^{36}$ erg s$^{-1}$) \dotfill &  2.95 & 3.16 & 3.23 & 3.23 \\
		H$_{\text{clump}}$  ($\times 10^{13}$ cm$^{-3}$) \dotfill & 7.94$^{+1.38}_{-1.63}$ & 3.54$^{+0.91}_{-1.03}$ & 3.46$^{+1.10}_{-0.95}$ & 1.00$^{+0.41}_{-0.68}$ \\
		H$_{\text{diffuse}}$ ($\times 10^{11}$ cm$^{-3}$) \dotfill & 1.41$\pm 0.10$ & 1.77$^{+0.17}_{-0.08}$ & 2.23$^{+0.21}_{-0.19}$ & 6.07$^{+1.23}_{-0.91}$ \\
		$\alpha$ $^a$ \dotfill &-3.0 & -3.0  & -3.0 &-3.0  \\
		R$_{in}$ $^b$ ($\times 10^{13}$ cm) \dotfill & 7.94 & 8.91 & 10.47 & 12.30\\
		Filling factor$^a$ \dotfill & 0.1 & 0.1 & 0.1 & 0.1 \\
		Power$^a$ \dotfill & 0.0 & 0.0 &0.0 &0.0 \\
		Covering factor (diffuse)\dotfill & 0.59 & 0.64 & 0.42 & 0.15\\
		Covering factor (clump)\dotfill & 0.41 & 0.36 & 0.58 & 0.55 \\
		He$^c$ \dotfill & 1.2 & 1.2 & 1.0 & 1.0 \\
		C$^c$ \dotfill & 13.5$^{+5.25}_{-3.00}$ & 20$^{+7.50}_{-3.50}$ & 15$^{+4.00}_{-2.00}$ & 8${\pm2}$\\
		N$^c$ \dotfill & 250 & 250 & 250 & 250\\
		O$^c$ \dotfill & 27$^{+3.00}_{-11.0}$ & 35$^{+5.25}_{-7.00}$ & 33$^{+5.00}_{-5.50}$ & 28${\pm5.25}$\\
		Na$^c$ \dotfill & 1.4$^{+0.20}_{-0.60}$ & 1.3$^{+0.40}_{-0.80}$ & 1.3$^{+0.40}_{-0.80}$ & 1.3$^{+0.40}_{-0.80}$\\
		Mg$^c$ \dotfill & 1.2$^{+0.49}_{-0.70}$ & 1.3$^{+0.50}_{-0.60}$ & 0.5$\pm 0.25$ & 0.5$\pm0.25$\\
		Fe$^c$ \dotfill & 1.0 & 1.0 & 1.0 & 1.0\\
		Carbon abundance in dust \dotfill &  \nodata & \nodata &  1.25 & 1.27  \\
		Silicate abundance in dust \dotfill  & \nodata & \nodata &  0.85 & 0.85 \\
		Total number of lines ($n$) \dotfill & 34 & 53 & 28 & 25\\
		Number of free parameters ($n_p$) \dotfill & 13 & 13 & 15 & 15\\
		Degrees of freedom \dotfill & 21 & 40 & 13 & 10 \\
		Total $\chi^2$ \dotfill & 40.01  & 78.10 & 23.82 & 16.79 \\
		Reduced $\chi^2$ \dotfill & 1.90 & 1.95 & 1.83 & 1.67\\
		\hline
	\end{tabular}}
		\end{center}
	\textbf{Notes:}\\
	$^a${This was not a free parameter in the model.}\\
	$^b${Initial radius of the ejecta. Calculated assuming an expansion velocity of 500 km s$^{-1}$ over 20 days, 23 days, 27 days, and 31 days after the outburst for epoch 1, epoch 2, epoch 3, and epoch 4, respectively. This was not a free parameter in the models.}\\
	$^c${The log abundance by number relative to hydrogen, relative to solar. All other elements that are not listed in the table were set to their solar values.}\\
\end{table*}

\begin{deluxetable*}{lllllllllllllllllllll}
\setlength{\tabcolsep}{1pt}
\tablecaption{Observed and best-fit CLOUDY model line fluxes for V1280 Sco \label{tab:flux}}
\tabletypesize{\scriptsize}
\tablehead{
\colhead{} & \colhead{} &
\colhead{} & \colhead{Epoch-01} & \colhead{} &
\colhead{} & \colhead{Epoch-02} & \colhead{} &
\colhead{} & \colhead{Epoch-03} & \colhead{} &
\colhead{} & \colhead{Epoch-04} & \colhead{} &\\
\colhead{LineID} & \colhead{$\lambda$ ($\mu$m)} &
\colhead{Observed Flux} & \colhead{Model Flux} & \colhead{$\chi^2$} &
\colhead{Observed Flux} & \colhead{Model Flux} & \colhead{$\chi^2$} &
\colhead{Observed Flux} & \colhead{Model Flux} & \colhead{$\chi^2$} &
\colhead{Observed Flux} & \colhead{Model Flux} & \colhead{$\chi^2$} &
}
\startdata
\hline
\multicolumn{14}{c}{Optical}\\\hline
H $\delta$  &  0.4102  &  \nodata &  \nodata  &  \nodata & 0.39E+00 & 0.20E-01 & 0.15E+01   &    \nodata  &  \nodata  &   \nodata  & \nodata & \nodata & \nodata \\
Fe II      &   0.4178 &  \nodata &  \nodata  &  \nodata & 0.23E+00 &  0.15E+00 &  0.71E-01   &    \nodata  &  \nodata  &   \nodata  & \nodata & \nodata & \nodata \\
Fe II      &   0.4233 &   \nodata &  \nodata  &  \nodata & 0.19E+00 &  0.22E+00 &  0.10E-01  &    \nodata  &  \nodata  &   \nodata  & \nodata & \nodata & \nodata \\
Fe II      &   0.4273  &      \nodata &  \nodata  &  \nodata & 0.14E+00 &  0.18E+00 &  0.17E-01 &    \nodata  &  \nodata  &   \nodata  & \nodata & \nodata & \nodata \\
H $\gamma$  &  0.4341 &  \nodata &  \nodata  &  \nodata & 0.49E+00 &  0.60E+00 &  0.13E+00   &    \nodata  &  \nodata  &   \nodata  & \nodata & \nodata & \nodata \\
Fe II      &   0.4417  &    \nodata &  \nodata  &  \nodata & 0.16E+00 &  0.14E+00 &  0.44E-02  &    \nodata  &  \nodata  &   \nodata  & \nodata & \nodata & \nodata \\
Fe II      &   0.4584  &   \nodata &  \nodata  &  \nodata & 0.21E+00 &  0.40E+00 &  0.40E+00  &    \nodata  &  \nodata  &   \nodata  & \nodata & \nodata & \nodata \\
Fe II      &   0.4629  &   \nodata &  \nodata  &  \nodata & 0.20E+00 &  0.25E+00 &  0.27E-01  &    \nodata  &  \nodata  &   \nodata  & \nodata & \nodata & \nodata \\
H $\beta$  & 0.4861 &   \nodata &  \nodata  &  \nodata &0.10E+01 &  0.10E+01 &  0.00E+00  &    \nodata  &  \nodata  &   \nodata  & \nodata & \nodata & \nodata \\
Fe II      &   0.4924  &   \nodata &  \nodata  &  \nodata &  0.34E+00 &  0.32E+00 &  0.44E-02   &    \nodata  &  \nodata  &   \nodata  & \nodata & \nodata & \nodata \\
Fe II      &   0.5018  &   \nodata &  \nodata  &  \nodata & 0.41E+00 &  0.60E+00 &  0.44E+00   &    \nodata  &  \nodata  &   \nodata  & \nodata & \nodata & \nodata \\
Fe II      &   0.5169  &   \nodata &  \nodata  &  \nodata & 0.47E+00  & 0.71E+00  & 0.93E+00  &    \nodata  &  \nodata  &   \nodata  & \nodata & \nodata & \nodata \\
Fe II      &   0.5234  &   \nodata &  \nodata  &  \nodata & 0.11E+00  & 0.15E+00 &  0.17E-01  &    \nodata  &  \nodata  &   \nodata  & \nodata & \nodata & \nodata \\
Fe II      &   0.5276  &  \nodata &  \nodata  &  \nodata & 0.12E+00 &  0.46E+00 &  0.12E+01  &    \nodata  &  \nodata  &   \nodata  & \nodata & \nodata & \nodata \\
Fe II      &   0.5317  &   \nodata &  \nodata  &  \nodata & 0.17E+00 &  0.21E+00 &  0.17E-01   &    \nodata  &  \nodata  &   \nodata  & \nodata & \nodata & \nodata \\
Fe II      &   0.5425  &   \nodata &  \nodata  &  \nodata & 0.50E-01 &  0.20E+00 &  0.25E+00  &    \nodata  &  \nodata  &   \nodata  & \nodata & \nodata & \nodata \\
Na I &D2, D1 &   \nodata &  \nodata  &  \nodata &  0.15E+00  & 0.26E+00 &  0.13E+00  &    \nodata  &  \nodata  &   \nodata  & \nodata & \nodata & \nodata \\
Fe II &0.6248  &   \nodata &  \nodata  &  \nodata & 0.80E-01 &  0.19E+00 &  0.13E+00 &    \nodata  &  \nodata  &   \nodata  & \nodata & \nodata & \nodata \\
Fe II & 0.6456  &   \nodata &  \nodata  &  \nodata & 0.80E-01 &  0.13E+00 &  0.27E-01  &    \nodata  &  \nodata  &   \nodata  & \nodata & \nodata & \nodata \\
H$\alpha$ & 0.6563  &   \nodata &  \nodata  &  \nodata & 0.28E+01 &  0.51E+01 &  0.58E+02  &    \nodata  &  \nodata  &   \nodata  & \nodata & \nodata & \nodata \\
\hline
\multicolumn{14}{c}{J-Band}\\\hline
				Pa $\gamma$& 1.0938  & 0.14E+01 & 0.45E+00 & 0.10E+02 & 0.11E+01 & 0.46E+00 & 0.49E+01 & 0.73E+00 &  0.25E+00 &  0.25E+01 & 0.66E+00 & 0.20E+00 &  0.23E+01 \\
				u.i     &    1.1126  & 0.50E-01 & 0.60E-01 & 0.11E-02 &0.30E-01  & 0.70E-01 &  0.17E-01 & 0.30E-01 &  0.30E-01 &  0.00E+00& 0.60E-01 &  0.40E-01 &  0.44E-02 \\
				O I     &    1.1287  & 0.17E+01 & 0.99E+00 & 0.60E+01& 0.14E+01 &  0.12E+01 &  0.28E+00&0.12E+01 &  0.12E+01 &  0.11E-02 & 0.10E+01 & 0.55E+00  & 0.23E+01 \\
				Na I    &    1.1404  & 0.6E+00 & 0.12E+00 & 0.26E+01&0.42E+00 &  0.17E+00 &  0.69E+00 & 0.58E+00 &  0.10E+00 &  0.25E+01& 0.16E+00 & 0.20E-01 & 0.21E+00 \\
				C I     &    1.1659  & 0.10E+01 & 0.11E+01 & 0.16E+00&0.75E+00 &  0.75E+00  & 0.00E+00 & 0.27E+00  & 0.10E+00 &  0.32E+00 & 0.20E+00 & 0.20E+00 & 0.00E+00 \\
				C I     &    1.1755  & 0.13E+01 & 0.82E+00 &  0.34E+01&0.11E+01 &  0.72E+00  & 0.19E+01 & 0.55E+00 &  0.56E+00 &  0.11E-02& 0.36E+00 & 0.20E+00 & 0.28E+00 \\
				Mg I    &    1.1828  & 0.12E+01 &  0.77E+00 &  0.23E+01&0.61E+00 &  0.43E+00  & 0.36E+00 &0.16E+00 &  0.13E+00  & 0.10E-01 &0.10E+00 &  0.37E+00 &  0.81E+00 \\
				N I     &    1.2469  & 0.13E+00 &  0.40E-01 &  0.90E-01& 0.90E-01 &  0.14E+00 &  0.27E-01& 0.40E-01 &  0.70E-01  & 0.10E-01 & 0.50E-01 & 0.50E-01 &  0.00E+00 \\
				C I     &    1.2562  &  0.17E+00 & 0.21E+00 &  0.17E-01& 0.22E+00 &  0.23E+00  & 0.11E-02 & 0.80E-01 &  0.23E+00  & 0.25E+00 & 0.90E-01 & 0.17E+00 &  0.71E-01 \\
				C I     &    1.2569  &  0.17E+00 & 0.14E+00 &  0.10E-01& 0.12E+00 &  0.23E+00  & 0.13E+00 & \nodata  & \nodata & \nodata & \nodata & \nodata & \nodata\\
				Pa $\beta$&  1.2818  & 0.10E+01 & 0.10E+01 &0.00E+00& 0.10E+01 &  0.10E+01 &  0.00E+00 & 0.10E+01 & 0.10E+01 &0.00E+00& 0.10E+01 &  0.10E+01 &  0.00E+00  \\
				O I     &    1.3164  & 0.17E+00 & 0.14E+00 & 0.10E-01 & 0.12E+00 &  0.22E+00 &  0.11E+00 & 0.40E-01 & 0.12E+00 &  0.71E-01 & 0.60E-01 & 0.11E+00 &  0.27E-01 \\				
\hline
\multicolumn{14}{c}{H-Band}\\\hline
				Mg I    &       1.5040  &   0.26E+00 &  0.20E+00  & 0.40E-01 &0.26E+00 &  0.47E+00  & 0.49E+00  & \nodata  & \nodata & \nodata & \nodata & \nodata & \nodata \\
     			Br 19   &       1.5256  &   0.90E-01  &0.80E-01 & 0.10E-2 & 0.80E-01 &  0.10E+00 &  0.44E-02 & 0.30E-01 &  0.17E+00  & 0.21E+00& 0.18E+00 & 0.70E-01 & 0.13E+00 \\
				Br 18   &       1.5341  &    0.16E+00 &  0.15E+00  & 0.11E-02& 0.16E+00  & 0.22E+00 &  0.40E-01 &0.17E+00 &  0.18E+00 &  0.11E-02 & \nodata  & 0.25E+00 & \nodata  \\
				Br 17   &       1.5439  &    0.27E+00  & 0.12E+00  & 0.25E+00&0.19E+00 &  0.14E+00 &  0.27E-01 & 0.36E+00 &  0.37E+00 &  0.11E-02 & 0.20E+00 & 0.21E+00 & 0.11E-02  \\
				Br 16   &       1.5557  &    0.39E+00  & 0.19E+00 &  0.44E+00& 0.35E+00  & 0.24E+00 &  0.13E+00& 0.52E+00 &  0.42E+00 &  0.11E+00 &  0.20E+00 &  0.25E+00 & 0.27E-01\\
				Br 14   &       1.5881  &    0.60E+00 &  0.58E+00 &  0.44E-02&0.50E+00 &  0.43E+00 &  0.54E-01 &0.62E+00 &  0.42E+00 &  0.44E+00 & 0.16E+00 & 0.27E+00 & 0.13E+00 \\
				C I     &       1.6005  &    0.20E+00 &  0.26E+00 &  0.40E-01&0.50E+00  & 0.52E+00 &  0.44E-02 & 0.18E+00 &  0.12E+00 &  0.40E-01 &\nodata  & \nodata & \nodata\\
				Br 13   &       1.6109  &   0.22E+00 &  0.13E+00 &  0.90E-01&0.32E+00 &  0.21E+00  & 0.13E+00 & 0.38E+00 &  0.30E+00 &  0.71E-01 & 0.31E+00 & 0.30E+00 &  0.11E-02 \\
				Br 12   &       1.6407  &   0.10E+01 &  0.10E+01 &  0.00E+00& 0.10E+01 &  0.10E+01  & 0.00E+00& 0.10E+01 &  0.10E+01 &  0.00E+00 & 0.10E+01 &  0.10E+01  & 0.00E+00 \\
				Br 11   &       1.6806  &   0.11E+01 &  0.33E+00 & 0.72E+01& 0.67E+00 &  0.55E+00  & 0.16E+00 & 0.17E+01 &  0.61E+00 &  0.13E+02 &  0.70E+00 & 0.27E+00 & 0.20E+01  \\
				C I     &       1.6890  &   0.12E+01 &  0.66E+00  & 0.33E+01&0.11E+01 &  0.97E+00 &  0.18E+00 & 0.18E+01 &  0.14E+01  & 0.11E+01& 0.17E+01 & 0.10E+01 & 0.48E+01 \\
				C I     &       1.7045  &    0.80E-01 &  0.13E+00 &  0.27E-01&0.20E-01  & 0.23E+00 &  0.49E+00 &0.40E+01 & \nodata & \nodata & \nodata & \nodata \\
				Mg I    &       1.7109  &  0.90E-01 &  0.17E+00 &  0.71E-01& 0.60E-01  & 0.20E+00  & 0.21E+00 & 0.14E+00  & 0.23E+00 &  0.90E-01& 0.14E+00 & 0.14E+00 & 0.00E+00 \\
				Br 10   &       1.7362  &  0.16E+01 &  0.12E+01 &  0.18E+01& 0.95E+00 &  0.12E+01 &  0.12E+01& 0.23E+01 &  0.20E+01 &  0.12E+01 &  0.18E+01 & 0.13E+01 & 0.33E+01 \\
\hline
\multicolumn{14}{c}{K-Band}\\\hline
				He I    &       2.0585  &  0.14E+00 & 0.90E-01 & 0.27E-01 & 0.17E+00 &  0.14E+00 &  0.10E-01 &  0.50E-01 &  0.50E-01 &  0.00E+00 &\nodata  & \nodata & \nodata    \\
				C I     &       2.1023  &   0.25E+00  & 0.33E+00 &  0.71E-01& 0.26E+00 &  0.60E+00  & 0.12E+01&  0.50E-01 &  0.30E+00 &  0.69E+00 & 0.18E+00 & 0.12E+00 &  0.40E-01  \\
                  CI     &       2.1295 & 0.45E+00  & 0.31E+00 &  0.21E+00& 0.33E+00  & 0.48E+00  & 0.25E+00& 0.13E+00 &  0.30E+00 &  0.32E+00 & 0.70E-01 & 0.14E+00 &  0.54E-01 \\
                  Br $\gamma$    &       2.1655 &  0.10E+01 &  0.10E+01 &  0.00E+00& 0.10E+01 &  0.10E+01 &  0.00E+00 &0.10E+01 &  0.10E+01 &  0.00E+00& 0.10E+01 &  0.10E+01 &  0.00E+00 \\
                  NaI     &       2.2056 &   0.21E+00 &  0.17E+00  & 0.17E-01& 0.20E+00  & 0.23E+00  & 0.10E-01& 0.10E-01 & 0.40E-01 &  0.10E-01 &\nodata  & \nodata & \nodata \\
                  CI     &       2.2906 &  0.41E+00 &  0.43E+00 &  0.44E-02& 0.42E+00  & 0.70E+00 &  0.87E+00 & \nodata  & \nodata & \nodata & 0.15E+00 & 0.20E+00 & 0.27E-01 \\
                  CI     &       2.3130 &  0.25E+00 &  0.47E+00 & 0.53E+00& \nodata & \nodata &\nodata & \nodata & \nodata  & \nodata & \nodata & \nodata  & \nodata \\
				\hline
				Total $\chi^2$ &&&& 40.01 &&& 78.10 &&& 23.82 &&& 16.79 \\
				\hline
\enddata
\end{deluxetable*}

\section{Summary}\label{summary}
We have modeled the optical and near-Infrared spectra of the dust forming nova V1280 Sco using the photoionization code \textsc{cloudy}, v.17.02. From the best-fit model, we estimate various physical and chemical parameters of the system and their changes with the formation of dust in the ejecta. We use a two-component model: a diffuse region with a low hydrogen density and a clump region with a high hydrogen density of ejecta to generate all emission lines present in the observed spectra. Our model predicts a very high value of ejecta density and high elemental abundances, especially in CNO. The enhanced values of elemental abundances and high ejecta density are consistent with a dust-forming nova arising on a CO white dwarf. We find that small amorphous carbon dust grains along with the large astrophysical silicate grains are present in the ejecta. The effects of dust during the post-dust phase are incorporated by creating dust opacities files using spherical Mie code in \textsc{cloudy}. Dust forms at the outer part of the ejecta where the density is high and the temperature is low. It appears that high ejecta density, low temperature, and high elemental abundances have favored dust nucleation in the nova. The weak line feature at 1.1126$\mu$m is identified as Fe II line from our model.

\section{Acknowledgements}
We would like to thank Prof. Steven N. Shore, the referee, for his valuable comments, suggestions and thought provoking questions on our paper which helped to improve the manuscript. The research work at S. N. Bose National Centre for Basic Sciences is funded by the Department of Science and Technology, Government of India. Gargi Shaw acknowledges WOS-A grant from the Department of Science and Technology (SR/WOS-A/PM-9/2017), Government of India.

\end{document}